\documentclass[aps,pra,twocolumn,showpacs]{revtex4-1}

\usepackage{amsmath,amssymb}
\usepackage[pdftex]{graphicx}

\usepackage{color}

\newcommand{\ii}{\mathrm{i}}

\let \Re \relax
\DeclareMathOperator{\Re}{Re}
\let \Im \relax
\DeclareMathOperator{\Im}{Im}

\begin{document}

\title{Collapse-revival dynamics and atom-field entanglement \\ in the non-resonant Dicke model}

\author{A. Alvermann}
\affiliation{Theory of Condensed Matter, Cavendish Laboratory, Cambridge CB3 0HE, United Kingdom}
\altaffiliation[Present address: ]{Institut f\"ur Physik, Ernst-Moritz-Arndt-Universit\"at, 17487 Greifswald, Germany}
\author{L. Bakemeier}
\author{H. Fehske}
\affiliation{Institut f\"ur Physik, Ernst-Moritz-Arndt-Universit\"at, 17487 Greifswald, Germany}

\begin{abstract}
We consider the dynamics of atomic and field coherent states in the non-resonant Dicke model.
At weak coupling an initial product state evolves into a superposition of multiple field coherent states that are correlated with the atomic configuration.
This process is accompanied by the buildup and decay of atom-field entanglement
and leads to the periodic collapse and revival of Rabi oscillations.
We provide a perturbative derivation of the underlying dynamical mechanism 
that complements the rotating wave approximation at resonance.
The identification of two different time scales explains how the 
dynamical signatures depend on the sign of detuning between the atomic and field frequency,
and predicts the generation of  either atomic or field cat states in the two opposite cases.
We finally discuss the restrictions that the buildup of atom-field entanglement during the collapse of Rabi oscillations imposes on the validity of semi-classical approximations that neglect entanglement.
\end{abstract}


\maketitle

\section{Introduction}

In cavity quantum electrodynamics~\cite{WVEB06},
where the confinement of atoms results in coherent coupling to a single field mode, optical signatures such as non-Poissonian photon statistics~\cite{Ca85,*RSW90} or vacuum Rabi oscillations~\cite{VBWW00}
allow for the direct observation of quantum effects on light-matter interaction.
A fundamental consequence of field quantization 
is the collapse and revival (CR) of Rabi oscillations in a resonant cavity~\cite{RWK87,*BSMDHRH96}.
This effect, described by the Jaynes-Cummings model~\cite{JC63,*WM08},
involves the generation of atom-field entanglement
and non-classical ``Schr\"odinger cat'' states of the photon field~\cite{ENSM80,*Gea90}.
In non-resonant cavities, on the other hand,
the preparation of field cat states relies on linear frequency shifts induced by the atom-field coupling~\cite{BHRDZ92,*BHDMMWRH96}.
Additional effects arise in 
situations beyond weak coupling or resonance~\cite{WSBFHMKGS04,*CBSNHM04},
but also for the superradiant quantum phase transition in atomic ensembles~\cite{BGBE10} which has a close connection to quantum chaos and critical entanglement~\cite{EB03,*LEB04}.

In this paper we analyze the CR dynamics of atomic ensembles in the non-resonant Dicke model~\cite{Di54}. We identify the relevant CR mechanism
that results from the weak-coupling correction to the bare atomic and field frequency.
In difference to the behavior of a single atom studied in the Jaynes-Cummings model, more complex CR patterns are expected for atomic ensembles~\cite{KS93,*MLRDR06,*JRGSSA10}. So far, they have been discussed only in the rotating wave approximation (RWA)~\cite{APS97,KS98},
which is restricted to the near-resonant case. 

The analysis of the non-resonant case provided here shows the importance of two
different time scales for the CR dynamics.
They are associated with the dynamical splitting of either field or atomic coherent states,
and the subsequent generation of field or atomic Schr\"odinger cat states.
The effectiveness of the different dynamical mechanisms depends
on the sign of detuning: Atomic (field) cat states are generated predominantly if the field frequency is larger (smaller) than the atomic transition frequency.
The former (latter) situation is further characterized by the buildup
of significant atomic squeezing (atom-field entanglement) during the initial collapse phase.
In both cases a periodic CR pattern develops on long time scales.

To understand these effects we proceed as follows.
We first describe, in Sec.~\ref{sec:CR}, the principal behavior using numerical results for 
atomic expectation values, the entanglement entropy and phase space distribution functions.
These results establish Rabi oscillations and CR dynamics for the non-resonant case and indicate  the evolution of an initial product state into a quantum superposition
with large atom-field entanglement.
We then deduce this behavior from the non-resonant weak-coupling perturbation theory developed in Sec.~\ref{sec:Pert} and App.~\ref{sec:app} as the equivalent to the RWA at resonance. 
Perturbation theory allows for a clear identification of the relevant mechanisms and predicts the characteristic structure of the wave function
as a quantum superposition of multiple classical field states.
In Sec.~\ref{sec:Cat} we return to numerical calculations for the opposite cases of small or large field frequency. According to the two different time scales that we found in perturbation theory
we will observe the realization of ``Schr\"odinger cat''-like states of the atom or the field.
Since atom-field entanglement plays a decisive role in the CR dynamics,
we discuss in Sec.~\ref{sec:SCA} the consequences for the standard semi-classical approximation
to the Dicke model, which is found to encounter problems even close to the classical field limit.
Our findings are summarized in Sec.~\ref{sec:Conc}.

\section{Collapse and revival dynamics}
\label{sec:CR}

Atomic ensembles are described in the Dicke model~\cite{Di54}
\begin{equation}\label{Ham}
 H  = - \Delta J_z + \lambda (a^\dagger + a) J_x + \Omega a^\dagger a \;,
\end{equation}
as a pseudo-spin of length $j$ (using angular momentum operators $J_x$, $J_z$), which represents an ensemble of $2j$ two-level atoms with transition frequency $\Delta$
coupled to a bosonic field mode with frequency $\Omega$ (using ladder operators $a^{(\dagger)}$).
We consider this model in the two different non-resonant cases $\Omega \ll \Delta$ and $\Omega \gg \Delta$.

\begin{figure}
\includegraphics[width=0.32\linewidth]{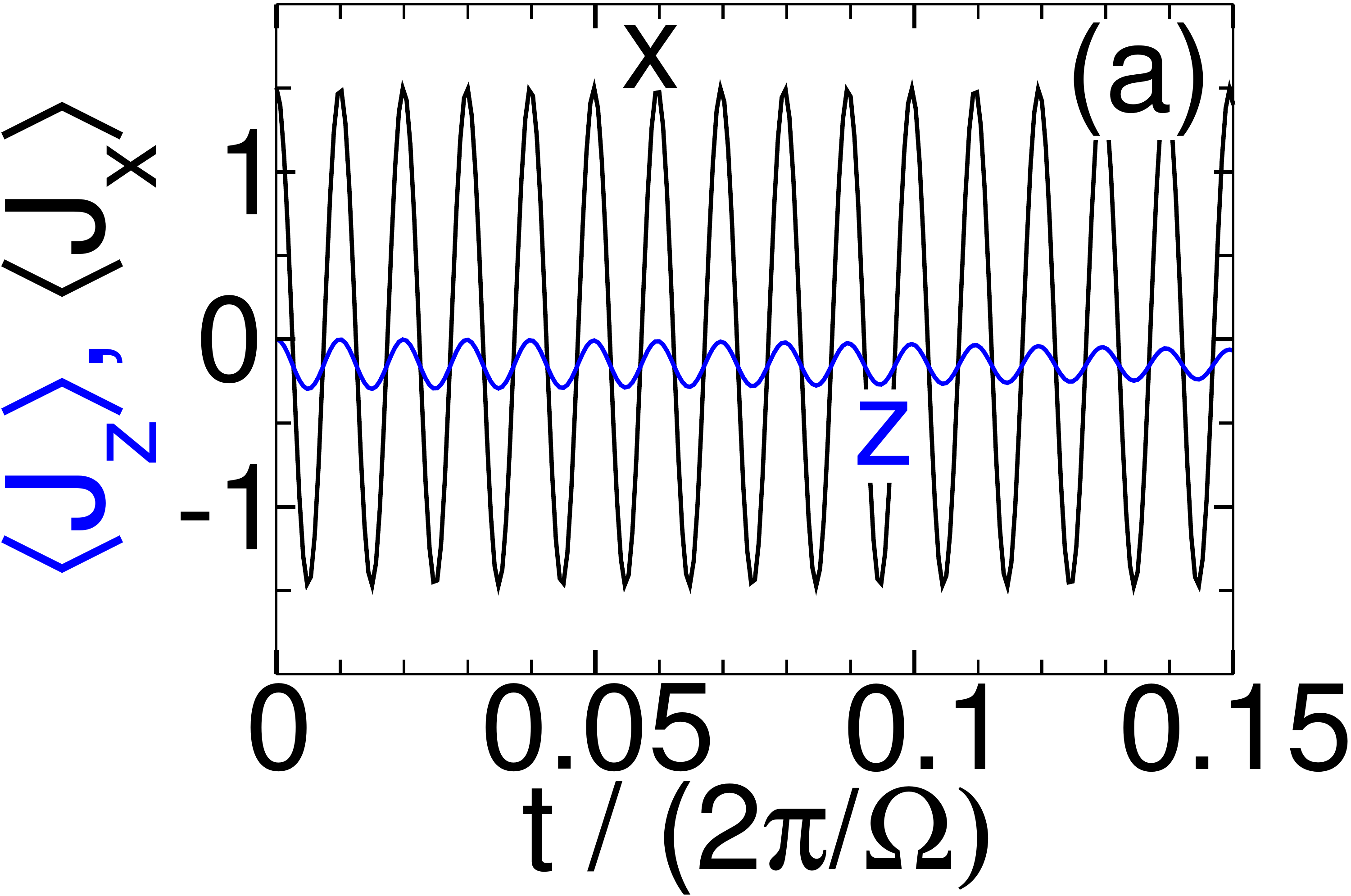} \hfill
\includegraphics[width=0.32\linewidth]{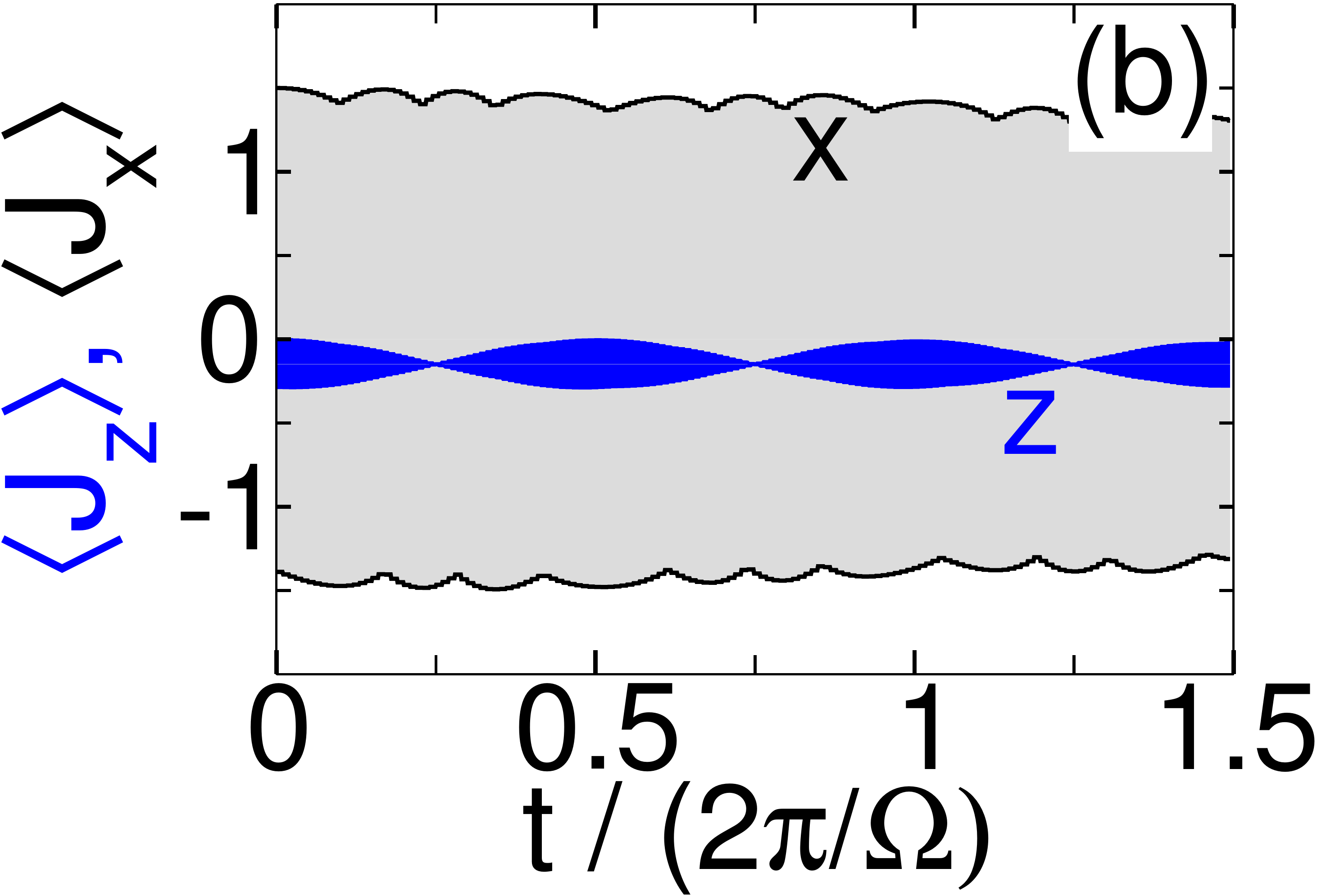} \hfill
\includegraphics[width=0.32\linewidth]{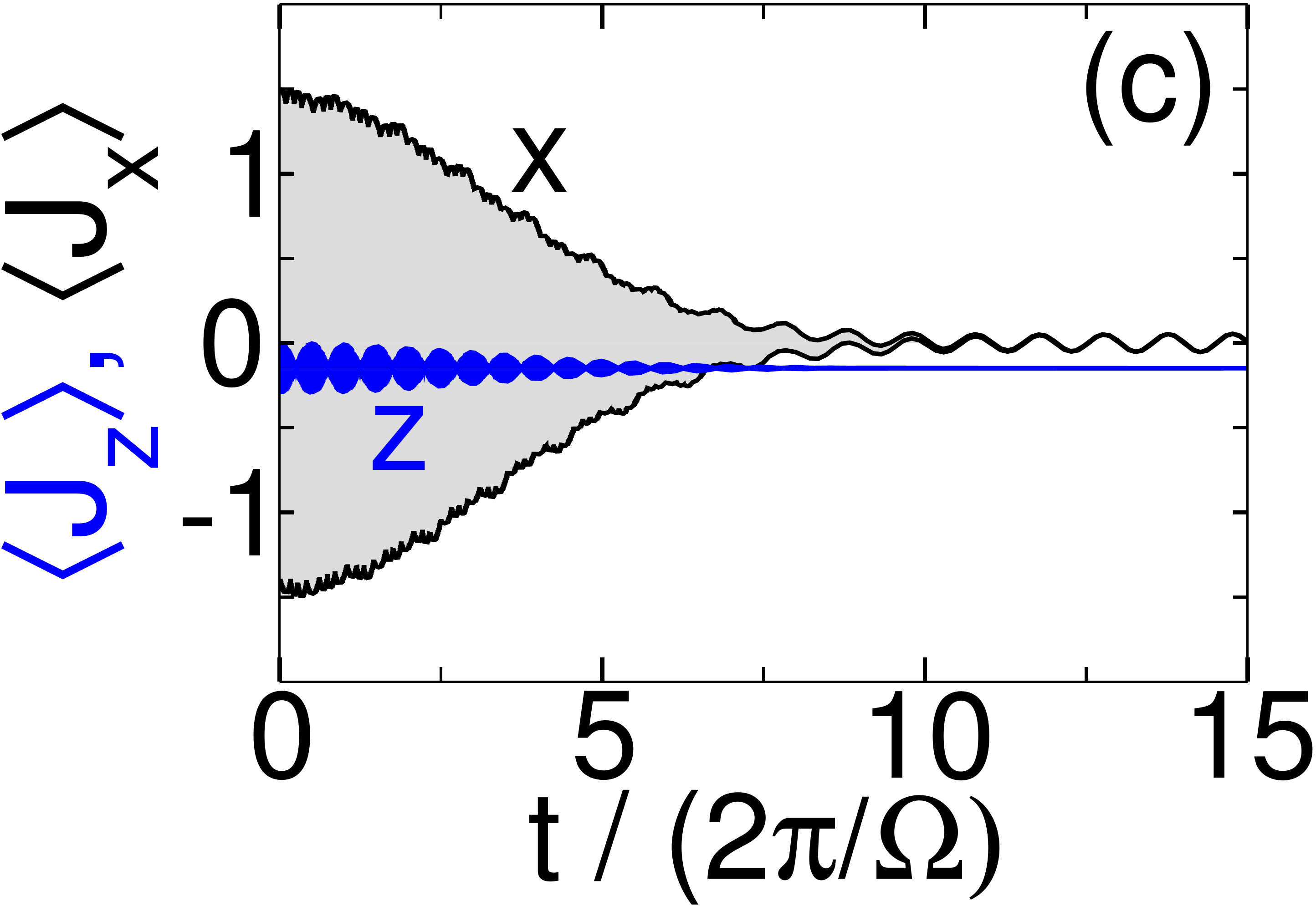} \hfill \\[0.5ex]
\includegraphics[width=0.48\linewidth]{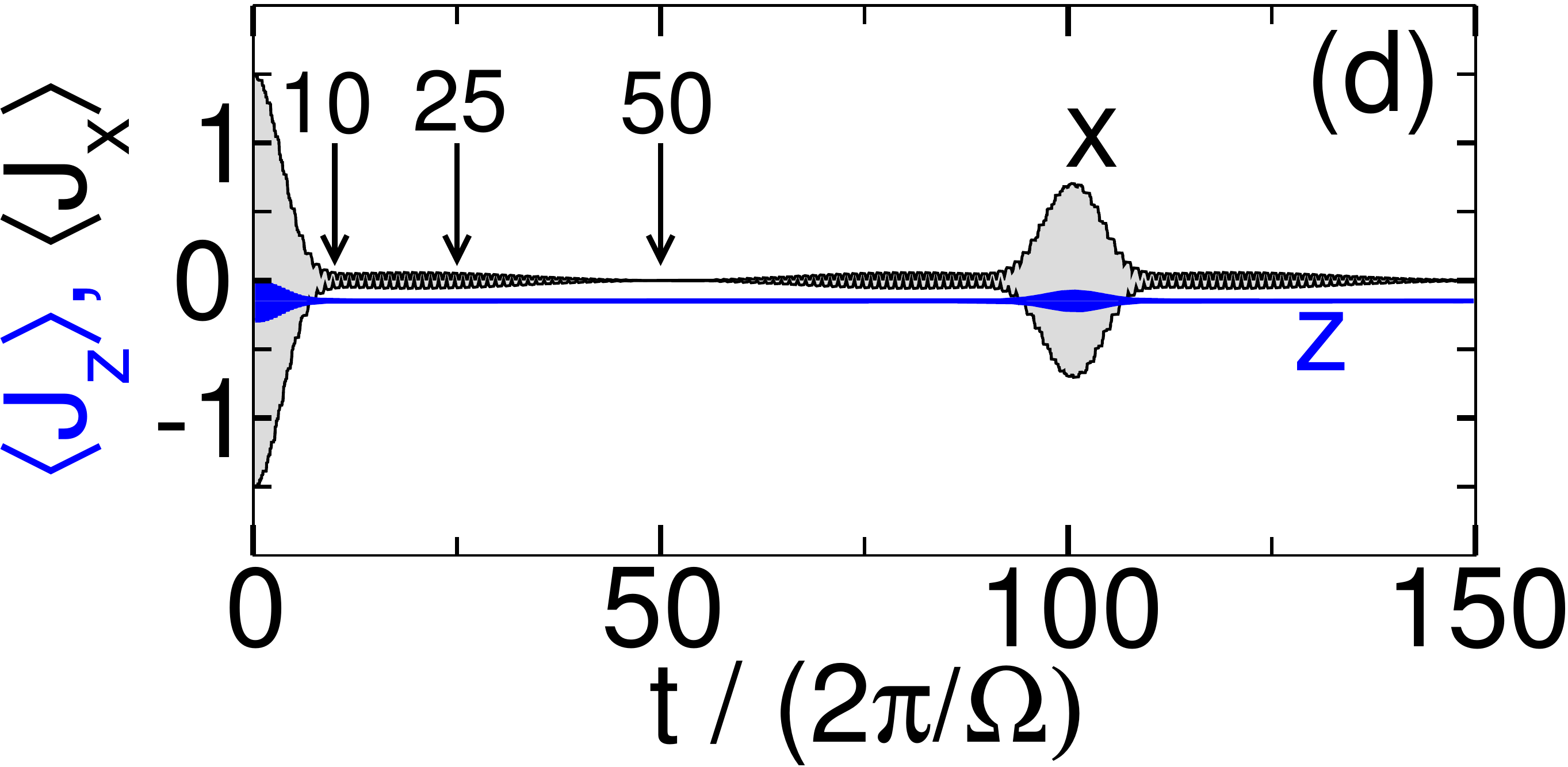} \hfill
\includegraphics[width=0.48\linewidth]{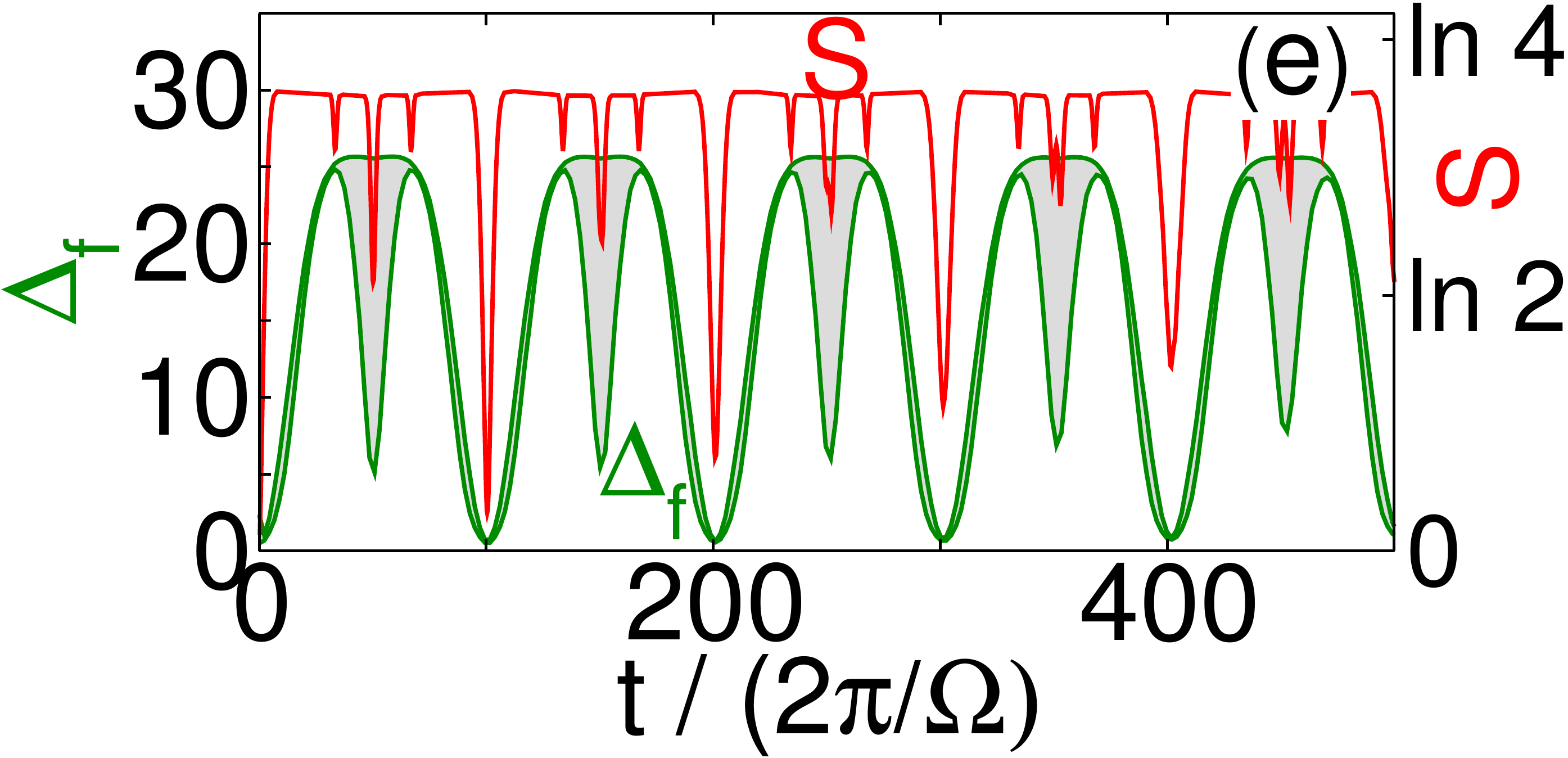}
\caption{(Color online)
CR dynamics of three atoms ($j=3/2$) for 
$\Omega/\Delta = 0.01$, $\lambda / \Delta = 0.01$,
with  $\theta_0 = \pi/2$, $\alpha_0=5.0$ in the initial state at $t=0$.
Upper row and lower left panel: Atomic expectation values $\langle J_x(t) \rangle$ (black curve), $\langle J_z(t) \rangle$ (blue curve) 
over different time scales, covering three orders of magnitude.
Panels (b)--(c) show envelope functions because the fast atomic oscillations visible in panel (a) can not be resolved in the pictures.
Lower right panel:
Entanglement entropy $S(t)$ (red curve) and field variance $\Delta_f(t)$ (green curve).
}
\label{fig:CR1}
\end{figure}

An example of CR dynamics for $\Omega \ll \Delta$ is given in Fig.~\ref{fig:CR1}. 
In this and all following examples,
the system at time $t=0$ is prepared in the product state
\begin{equation}
|\psi(0)\rangle = |\theta_0\rangle \otimes |\alpha_0\rangle
\end{equation}
of an atomic coherent state 
\begin{multline}\label{SpinCoh}
|\theta_0\rangle =  \exp[ - \ii \theta_0 J_y] |j,j \rangle \\
= \sum\limits_{m=-j}^j {\textstyle \binom{2j}{j+m}^{1/2}} (\cos \tfrac{\theta_0}{2})^{j+m}
 (\sin \tfrac{\theta_0}{2})^{j-m} |j,m\rangle
\end{multline}
and a field coherent state  
\begin{equation}
\begin{split}
|\alpha_0 \rangle &= \exp [\alpha_0 a^\dagger - \alpha_0^* a ] |0Ê\rangle \\
&=e^{-|\alpha_0|^2} \sum\limits_{n=0}^\infty \frac{\alpha_0^n}{\sqrt{n!}} |n\rangle
\;,
\end{split}
\end{equation}
using the $J_z$-eigenstates $|j,m\rangle$ and the $a^\dagger a$-eigenstates $|n\rangle$
(cf. Ref.~\cite{ZFG90}).
In the initial state it is $\langle J_z\rangle = j \cos \theta_0$,
$\langle J_x \rangle = j \sin \theta_0$ and
$\langle a \rangle = \alpha_0$. 
We assume that $|\alpha_0| \gg 1$ for a classical field, and generally choose $\alpha_0 \in \mathbb{R}$, $\alpha_0 > 0$.

Fig.~\ref{fig:CR1} shows the atomic observables $\langle J_{x/z}(t) \rangle$ over 
different time scales.
They have been calculated from numerical time-propagation of the wave function using the Chebyshev technique~\cite{TK84}.
Up to $10^3$ bosons have been kept in the calculations to prevent errors from the truncation of the bosonic part of the Hilbert space.
All numerical data shown are exact in the sense that the relative error is on the level of machine precision.

On a short time scale (panel (a)), we observe fast oscillations with the atomic frequency $\Delta$.
The amplitude of the $\langle J_z(t) \rangle$-oscillations is of order $\lambda$.
On a longer time scale (panel (b)), the atomic oscillations in $\langle J_z(t) \rangle$ are modulated by oscillations with frequency $\Omega$
(note that we show envelope functions whenever the fast atomic oscillations are not resolved in the pictures).
We call them Rabi oscillations in analogy to the resonant case because they arise from the coupling of the atoms to a classical field. In the non-resonant case $\Omega \ll \Delta$ they appear with the field frequency $\Omega$.
The collapse of Rabi oscillations is observable over the first 5--10 field periods $2\pi/\Omega$
(panel (c)), before they reappear on an even longer time scale (panel (d)),
with a revival time of $T_R / ( 2\pi/\Omega) \approx 100$ field oscillations. 
In contrast to the resonant case, a periodic CR pattern of Rabi oscillations evolves. 

The CRs are accompanied by the periodic buildup and decay of atom-field entanglement (lower right panel in Fig.~\ref{fig:CR1}), which we measure through the entanglement entropy 
\begin{equation}
S =  - \mathrm{Tr} [\rho_r \ln \rho_r]  \;.
\end{equation}
It is obtained from either the reduced atomic or field density matrix $\rho_r$, which both give the same value according to the Schmidt decomposition~\cite{HHHH09}.
From the initial product state with $S=0$,
entanglement is generated in the collapse phase until $S$ is close to the maximal possible value $\ln (2 j +1)$. The revivals coincide with entanglement decay, as the wave function returns to a product form. 
A similar behavior is found for the field variance, 
which we define as the product
\begin{equation}
\Delta_f = \frac{1}{2} \Big(\Delta [a^\dagger+a] \, \Delta [\ii(a^\dagger-a)] \Big)^{1/2}
\end{equation} 
of the uncertainties of the field operators $(a^\dagger + a)/\sqrt{2}$, $\ii (a^\dagger -a)/\sqrt{2}$,
where $\Delta A = \langle A^2 \rangle - \langle A \rangle^2$ as usual. 
In Fig.~\ref{fig:CR1} it signals that the initial coherent state with minimal $\Delta_f = 1/2$ evolves into a field state with large variance in the collapse phase.

\begin{figure}
\includegraphics[width=0.30\linewidth]{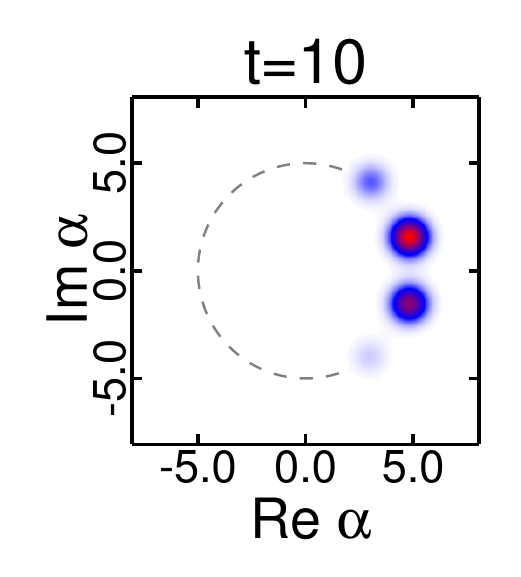}
\includegraphics[width=0.30\linewidth]{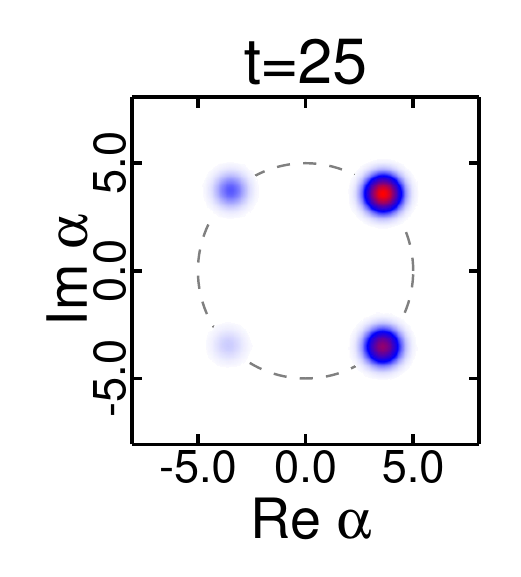} 
\includegraphics[width=0.37\linewidth]{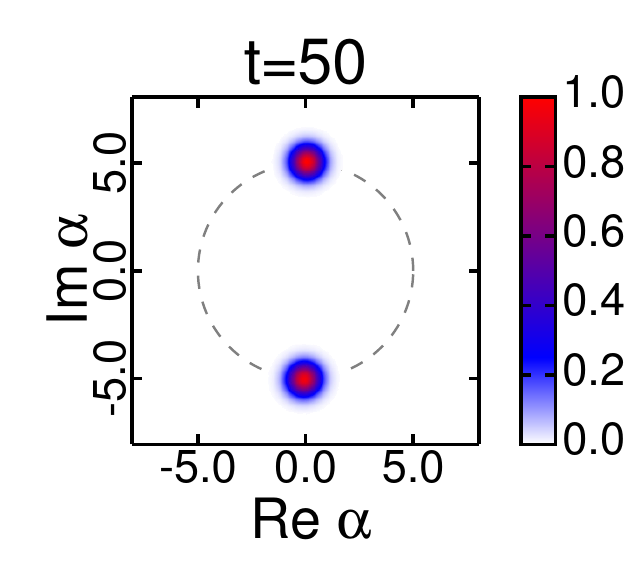}
\caption{(Color online)
Field Husimi function $Q_f(\alpha)$ 
in the collapse phase for three atoms with parameters from Fig.~\ref{fig:CR1},
at times $t/(2\pi/\Omega) = 10, 25, 50$ corresponding to the arrows in the lower left panel in Fig.~\ref{fig:CR1}.
The dashed circles have radius $\alpha_0$.
The color encoding runs from white for $Q=0$ through blue to red for the maximal $Q$-value in the respective picture, as shown in the color bar on the right.
}
\label{fig:CR1Hus}
\end{figure}

To identify the structure of the bosonic field we show in Fig.~\ref{fig:CR1Hus} the field Husimi function~\cite{ZFG90}
\begin{equation}\label{QField}
Q_f(\alpha) = |\langle \alpha |Ê\psi \rangle|^2 \;.
\end{equation}
By definition, it is the probability of finding the field in the coherent state $|\alpha\rangle$. 
For $\lambda=0$, points in the phase space rotate on circles around the origin $\alpha=0$.
We observe that the Husimi function is a superposition of several Gaussian peaks indicating the splitting of the initial into multiple coherent field states in the collapse phase.
Partial revivals, e.g. at $t=T_R/2 \approx 50 \times (2 \pi)/\Omega$, occur when some but not all of the coherent field states merge.
At a full revival $t=T_R$ (not shown), the Husimi function consists again of a single Gaussian peak at $\alpha_0$ which indicates the revival of the initial state.
This explains the behavior of the field variance $\Delta_f$.

We observe here the first of two CR mechanisms.
An initial coherent field state splits because the rotation frequency in oscillator phase space depends on the atomic configuration.
The field-induced collapse phase coincides with maximal entanglement 
between the atomic and several coherent field states whose phase space position differs by a finite angle. 
The large entanglement indicates an incoherent superposition rather than a field cat state that requires a coherent superposition~\cite{SPK91}.
Periodic revivals occur when the relative angles approach zero again.

\section{Non-resonant perturbation theory}
\label{sec:Pert}

After the discussion of the first example,
we now derive the CR mechanisms using perturbation theory.
The central result will be that, under the appropriate conditions for the perturbative treatment specified below,
the wave function has the structure
\begin{equation}\label{PsiStruct}
 |\psi(t)\rangle =  \sum_{m=-j}^j \psi_m(t) \; |\sigma_m(t)\rangle  \otimes |\alpha_m(t)\rangle \;,
\end{equation} 
where $|\sigma_m(t)\rangle$ are a set of atomic (i.e. spin) states 
that occur together with field coherent states $|\alpha_m(t)\rangle$.
We note that  the field remains essentially classical during time evolution since only coherent states occur, but it becomes entangled with the atomic ensemble  because of the dependence of $\alpha_m(t)$ on $m$:
Each classical field state is ``tagged'' by the associated atomic configuration $|\sigma_m(t)\rangle$.

\subsection{Derivation of the perturbative wave function}

 The unperturbed eigenstates at zero coupling ($\lambda=0$)
 are the product states $|m\rangle \otimes |n\rangle$
of $J_z$-eigenstates $|m\rangle$ and field Fock states $|n\rangle$,
with energy $E^0_{mn} = - m \Delta + n \Omega$.
From standard non-degenerate second order perturbation theory we obtain 
the energy correction as
\begin{equation}
  E^{(2)}_{mn} =  \frac{\lambda^2}{8}  \!\! \sum_{\mu \nu = \pm 1} \frac{ (j- \mu m)(j+\mu m+1)(2 n+\nu +1) }{\mu \Delta - \nu \Omega}  \;,
\end{equation}
where the sum contains contributions from the four states
$|m \pm 1 \rangle \otimes |n \pm 1 \rangle$ contributing in second order
through the interaction term $\lambda (a^\dagger + a) J_x $.
The correction to the eigenstates is given by a similar expression,
but we need to keep only the leading first order terms in $\lambda$ (further perturbative results are given in App.~\ref{sec:app}).
Both expressions can be combined into 
a unitary time-evolution operator
that gives the perturbative wave function as
\begin{equation}\label{PertWave}
|\psi(t)\rangle_{(2)} = U^\dagger \exp [ -\ii \tilde{H} t ] U |\psi(0)\rangle \;,
\end{equation}
with an effective Hamiltonian
\begin{equation}\label{Heff}
 \tilde{H} =  - \Delta J_z + \Omega a^\dagger a  - \omega_E (2 a^\dagger a +1) J_z - \omega_S J_z^2  
\end{equation}
and a unitary transformation of states 
\begin{equation}\label{Utraf}
  U  = \exp \Big[  - \frac{2 \omega_S}{\lambda} (a^\dagger -a ) J_x 
  -  \ii \frac{2 \omega_E}{\lambda} (a^\dagger + a ) J_y   \Big] \;.
\end{equation}
We here introduced the two frequencies  
\begin{equation}
 \omega_E = \frac{\lambda^2 \Delta}{2(\Delta^2-\Omega^2)} \;, \quad
 \omega_S = \frac{\lambda^2 \Omega}{2(\Delta^2-\Omega^2)}
\end{equation}
that appear in $\tilde{H}$ as a consequence of the atom-field coupling,
and dropped a constant term $\Omega j (j+1)$.
Due to the unitary form of the perturbative result it remains valid for long times $t$ and large $|\alpha_0|$,
provided that $\lambda |\alpha_0| \ll |\Delta^2 - \Omega^2|$.

The central information about the non-resonant CR mechanism 
is contained in the two time scales
\begin{equation}\label{TS}
 T_{E} =  \frac{\pi}{ |\omega_E|}
 \;, \quad
 T_{S} = \frac{\pi}{ |\omega_S|}
\end{equation}
in the effective Hamiltonian $\tilde{H}$.
The ``entangling'' time $T_E$
is associated with the term $a^\dagger a J_z$,
which gives an energy correction $\propto mn$ and
is the origin of the atom-dependent field splitting observed in Fig.~\ref{fig:CR1Hus}. 
The ``squeezing'' time $T_S$
occurs with $J_z^2$.
Since this term affects only the atomic ensemble, no additional entanglement is generated.
Instead, it leads to the squeezing of atomic coherent states and splitting into atomic cat states~\cite{KU93,APS97}.
The ratio $T_{E} / T_{S} = \Omega /  \Delta$ determines which term dominates the initial dynamics.
A similar perturbation theory for a simplified model in RWA does not distinguish between the different time scales~\cite{KS98} (see also App.~\ref{sec:app}).

We can now construct the perturbative wave function in the form of Eq.~\eqref{PsiStruct},
starting from the initial state $|\psi(0)\rangle = |\theta_0\rangle \otimes |\alpha_0 \rangle$ used  throughout this paper. 
Under the assumption $|\alpha_0| \gg 1$
we can replace the operators $a^\dagger + a$, $\ii (a^\dagger -a)$ in the unitary transformation $U$  from Eq.~\eqref{Utraf}
by the scalars $2 \Re \alpha_0$, $2 \Im \alpha_0$ respectively.
The error of this replacement is of order $1/|\alpha_0|$.
Then, $U$ reduces to a spin rotation operator of the form
\begin{equation}\label{ROp}
  R(a,b) = \exp \Big[ \ii (a J_x -  b J_y)  \Big] \qquad (a,b \in \mathbb{R}) \;,
\end{equation}
and we get
\begin{equation}\label{URotates}
 U \Big[ |\sigma\rangle \otimes |\alpha\rangle \Big] = \Big[ R\Big(\frac{4 \omega_S \Im \alpha}{\lambda}   ,  \frac{4 \omega_E \Re \alpha}{\lambda}  \Big)  |\sigma\rangle \Big] \otimes |\alpha\rangle
 \end{equation}
for every atomic state $|\sigma\rangle$ and a field coherent state with $|\alpha| \gg 1$.

We note that general atomic coherent states can be defined through
\begin{equation}\label{SpinCohGen}
 |\theta, \phi \rangle = R(\theta \sin \phi, \theta \cos \phi ) | j,j \rangle \;,
\end{equation}
which gives 
\begin{equation}\label{SpinCohExp}
\begin{split}
 \langle \theta, \phi| J_z | \theta, \phiÊ\rangle &= j \cos \theta \;, \\
 \langle \theta, \phi| J_x | \theta, \phiÊ\rangle &= j \sin \theta \cos \phi \;. 
\end{split}
\end{equation}
Atomic coherent states remain coherent states under rotation.
In particular for the initial state we have
\begin{equation}\label{UToInitial}
 U \Big[ |\theta_0\rangle \otimes |\alpha_0 \rangle \Big] = |{\theta_0 \! + \! \delta \theta }\rangle \otimes |\alpha_0 \rangle 
\end{equation}
from the relation $R(0,\theta')|\theta\rangle = |\theta+\theta'\rangle$,
a simple rotation of the atomic coherent state by the angle 
\begin{equation}
\delta \theta =   \frac{2  \lambda \Delta \alpha_0}{\Delta^2-\Omega^2} \;.
\end{equation}
Here, we still assume for simplicity that $\alpha_0 \in \mathbb{R}$.

Since the effective Hamiltonian $\tilde{H}$ is diagonal in the $J_z$-eigenstates $|m\rangle$,
application of $\exp[-\ii \tilde{H} t ]$ to the state in Eq.~\eqref{UToInitial} rotates the field component of the different $J_z$-contributions.
The operator $a^\dagger a$ generates a rotation of field coherent states of the form
$\exp[ \ii \xi a^\dagger a ] |\alpha\rangle = |{\exp(\ii \xi) \alpha}\rangle$,
such that we have
\begin{multline}\label{ExpIH}
 \exp[ - \ii \tilde{H} t ] |m\rangle \otimes |\alpha_0\rangle \\[0.5ex] = e^{\ii t \left( m (\Delta + \omega_E)  + m^2 \omega_s \right) } \; |m\rangleÊ\otimes  |\alpha_m(t)\rangle \;,
\end{multline}
with
\begin{equation}\label{AlphaM}
 \alpha_m (t) = \alpha_0 e^{-\ii t (\Omega - 2 m \omega_E  )}  \;.
\end{equation}
We note the scalar product 
\begin{equation}\label{AlphaScal}
| \langle \alpha_m(t) | \alpha_{m'}(t) \rangle |=  \exp \Big( \!  - |\alpha_0|^2 (1- \cos \delta \alpha_{mm'}   ) \Big)
\end{equation}
between two field coherent states enclosing the finite phase space angle
$\delta \alpha_{mm'} = 2 \pi (m-m')  (t /T_E)$.

For each of the states in Eq.~\eqref{ExpIH} the inverse transformation $U^\dagger$ leads again to a spin rotation as in Eq.~\eqref{URotates},
but the arguments of $R(\cdot,\cdot)$ now depend on $\alpha_m(t)$.
We have $U^\dagger |m\rangle \otimes |\alpha_m(t)\rangle = |\sigma_m(t) \rangle \otimes |\alpha_m(t)\rangle$
with
\begin{equation}\label{SigmaM}
 |\sigma_m(t)\rangle = R\Big(- \frac{4 \omega_S}{\lambda} \Im \alpha_m(t),
 -  \frac{4 \omega_E}{\lambda} \Re \alpha_m(t) \Big) |m\rangle \;.
\end{equation}

Collecting all results, we finally see that the perturbative wave function $|\psi(t)\rangle_{(2)}$ 
has indeed the structure proposed in Eq.~\eqref{PsiStruct},
with individual terms given by Eqs.~\eqref{AlphaM},~\eqref{SigmaM}
and
\begin{equation}\label{PsiM}
 \psi_m(t) =   e^{\ii t \left( m (\Delta + \omega_E)  + m^2 \omega_s \right) }  \, \langle m |{\theta_0 \! + \! \delta \theta} \rangle \;.
\end{equation}
The coefficients $ \langle m |\theta  \rangle$ of an atomic coherent state are given in Eq.~\eqref{SpinCoh}.
We note that the spin states $|\sigma_m(t)\rangle$ are not orthogonal since the rotation with $R(\cdot,\cdot)$ depends on $\alpha_m(t)$.
This effect is of order $\lambda$ and not an artifact of perturbation theory.

\subsection{Collapse and revivals in perturbation theory}

\begin{figure}
\includegraphics[width=0.32\linewidth]{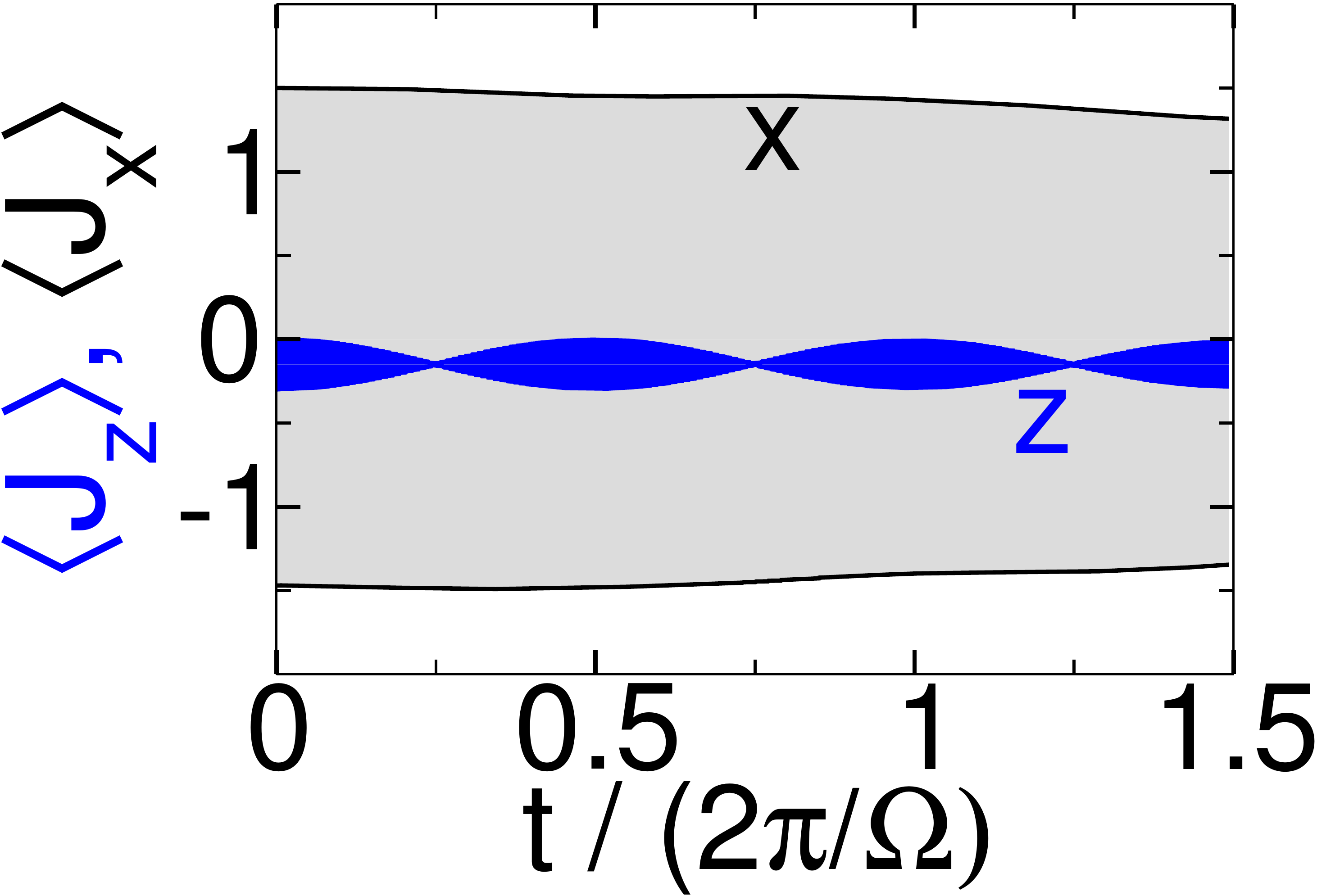} \hfill
\includegraphics[width=0.32\linewidth]{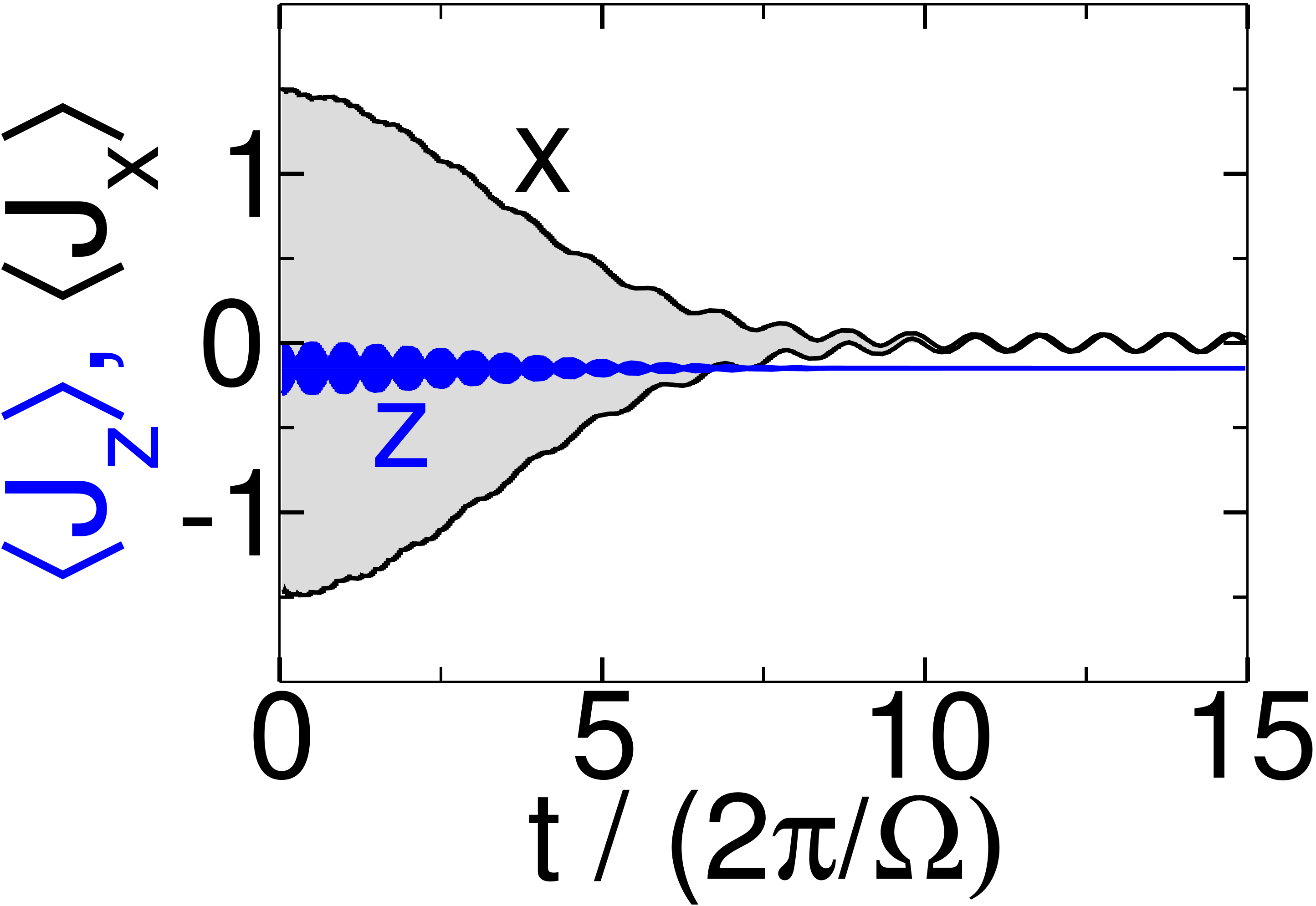} \hfill
\includegraphics[width=0.32\linewidth]{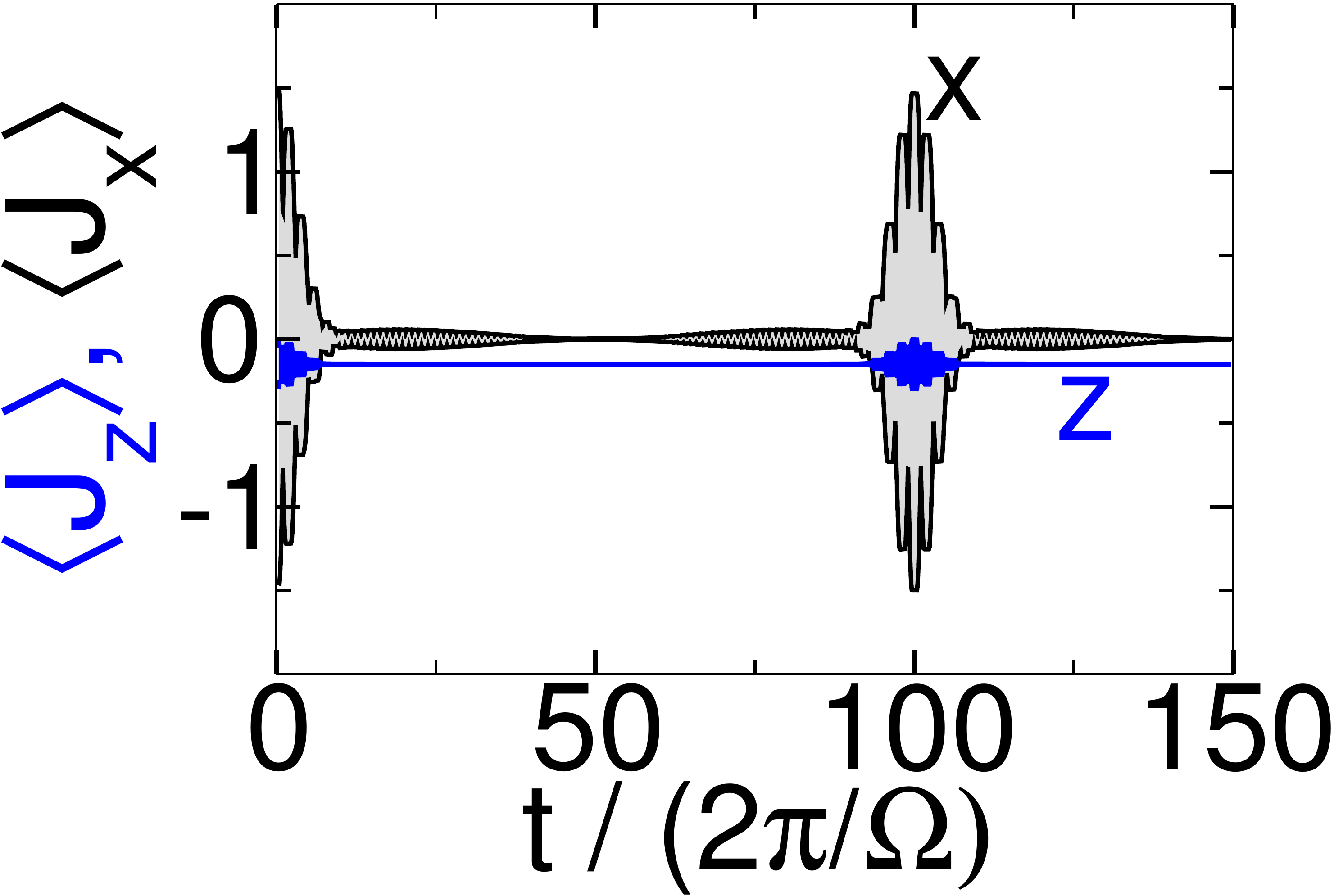} \hfill
\caption{(Color online)
Perturbative result for the CR dynamics of three atoms with parameters from Fig.~\ref{fig:CR1}.
Shown are the atomic expectation values $\langle J_x(t) \rangle$, $\langle J_z(t) \rangle$ obtained with Eqs.~\eqref{PsiStruct}--\eqref{PsiM}, corresponding to panels (b)--(d) in Fig.~\ref{fig:CR1}.
}
\label{fig:CR1Pert}
\end{figure}

Explicit expressions for the values of atomic and field observables can be derived from the above equations, but we do not show them here since they are rather lengthy and not very illuminating.
Instead, let us focus on the CR dynamics in the situation $\Omega \ll \Delta$ addressed in Figs.~\ref{fig:CR1},~\ref{fig:CR1Hus}.
The perturbative result for atomic expectation values is shown in Fig.~\ref{fig:CR1Pert}.
Comparison with Fig.~\ref{fig:CR1} shows that perturbation theory is in excellent agreement with the numerical data.

In the initial dynamics, for $t \ll T_E, T_S$, differences between the coherent state parameters $\alpha_m(t)$ are negligible.
Therefore, the wave function has product form $|\psi(t)\rangle \approx |\sigma(t)\rangle \otimes |\alpha_0 e^{-\ii \Omega t}\rangle$.
The atomic state $|\sigma(t)\rangle$ is obtained from the initial atomic coherent state  $|\theta_0\rangle$ through three rotations around different axes:
The first rotation from Eq.~\eqref{UToInitial},
the second rotation from the effective Hamiltonian as $\exp[\ii t \Delta  J_z]$,
and the third rotation from Eq.~\eqref{SigmaM} which presently does not depend on $m$.
Therefore, $|\sigma(t)\rangle$ is itself an atomic coherent state as in Eq.~\eqref{SpinCohGen}.
From Eq.~\eqref{SpinCohExp}
we obtain the atomic expectation values as
\begin{multline}\label{PertRabi}
 \langle J_z(t) \rangle = j \Big(\cos \Delta t \sin (\delta \theta \cos \Omega t)  \sin (\theta_0 + \delta \theta)  \\
 + \, \cos (\delta \theta \cos \Omega t) \cos (\theta_0 + \delta \theta)    \Big) \;,
\end{multline}
\begin{multline}
 \langle J_x(t) \rangle = j \Big(\cos \Delta t \cos (\delta \theta \cos \Omega t) \sin (\theta_0 + \delta \theta)  \\
 - \, \sin (\delta \theta \cos \Omega t) \cos (\theta_0 + \delta \theta)    \Big)  \;. 
\end{multline} 
These expressions describe, through the term $\delta \theta \cos \Omega t$, Rabi oscillations with frequency $\Omega$.
Their origin within perturbation theory is the dependence of the final rotation in Eq.~\eqref{SigmaM} on $\alpha_m(t) \approx \alpha_0 e^{-\ii \Omega t}$ for $t \ll T_E$.

Since $\Omega \ll \Delta$,
the term $a^\dagger a J_z$ determines the CR dynamics for times $t \sim T_E \ll T_S$.
The splitting of the initial coherent field state $|\alpha_0\rangle$ into the $2j+1$ coherent states 
$|\alpha_m\rangle$ is the source of entanglement with the atomic $J_z$-eigenstates $|m\rangle$.
Because different $|m\rangle$-states are orthogonal, the field is in an incoherent superposition.
The collapse of Rabi oscillations is a consequence of the decreasing overlap 
$| \langle \alpha_m |Ê\alpha_{m'} \rangle |$ from Eq.~\eqref{AlphaScal}.

Deep in the collapse phase
the overlap between different contributions $|\sigma_m\rangle \otimes |\alpha_m\rangle$ in Eq.~\eqref{PsiStruct} is negligible.
As a consequence the phase of $\psi_m(t)$ in Eq.~\eqref{PsiM}, and thus the term $\Delta t$ responsible for atomic oscillations, drops out of the expressions.
The atomic expectation values are now
\begin{equation}
 \langle J_z(t) \rangle = \sum_{m=-j}^j m \, |\langle m | \theta_0 + \delta \theta \rangle|^2  \cos \xi_m(t) \; ,
\end{equation}
\begin{equation}
 \langle J_x(t) \rangle = \sum_{m=-j}^j m \, |\langle m | \theta_0 + \delta \theta \rangle|^2 \sin \xi_m(t) \; , 
\end{equation}
with $\xi_m(t) = \delta \theta \cos (\Omega - 2 \omega_E m)t$.
Expanding $\cos \xi_m(t) = 1 + O(\lambda^2)$ we see that,
up to small corrections, $\langle J_z \rangle$ in the collapse phase is given by the constant value
\begin{multline}
 \langle J_z(t) \rangle|_\mathrm{collapse} = \sum_{m=-j}^j m \, |\langle m | \theta_0 + \delta \theta \rangle|^2 \\
 = \langle \theta_0 + \delta \theta | J_z | \theta_0 + \delta \theta \rangle = j \cos (\theta_0 + \delta \theta) \;.
\end{multline}
This result is again a consequence of the rotation of the initial atomic coherent state in Eq.~\eqref{UToInitial}.
For  the parameters in Fig.~\ref{fig:CR1}
the expected value is $j \cos (\theta_0 + \delta \theta) \approx - 0.15$, which is in full agreement with the numerical results.
On the other hand, $\langle J_x(t) \rangle$ retains through the $\sin \xi_m(t)$ term of order $\lambda$ the $\cos (\Omega t)$-dependence seen in the middle panel in Figs.~\ref{fig:CR1},~\ref{fig:CR1Pert}.

For later times, inspection of Eqs.~\eqref{AlphaM},~\eqref{AlphaScal} shows that periodic revivals occur at multiples of $T_E$. For Fig.~\ref{fig:CR1}, the estimate $T_E = 100 (2 \pi/\Omega)$ is close to the numerical value.
The appearance of a periodic CR pattern is a consequence of the linear dependence of 
$E_{mn}^{(2)}$ on both $m$, $n$,
which is a significant difference from the RWA at resonance, 
where the energy correction $\propto \sqrt{n}$ prevents truly periodic revivals~\cite{ENSM80,Gea90}.

\begin{figure}
\includegraphics[width=0.48\linewidth]{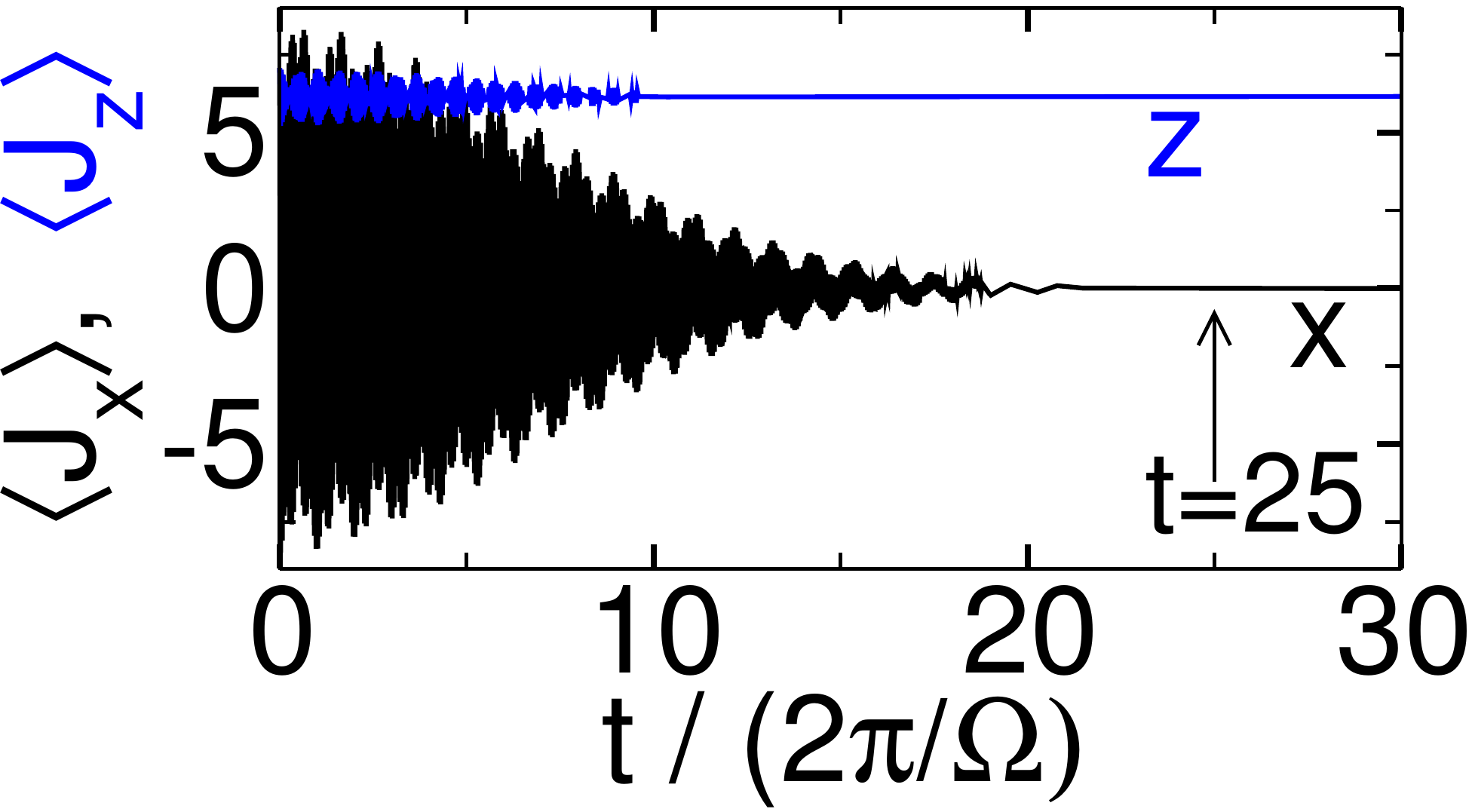} \hfill
\includegraphics[width=0.48\linewidth]{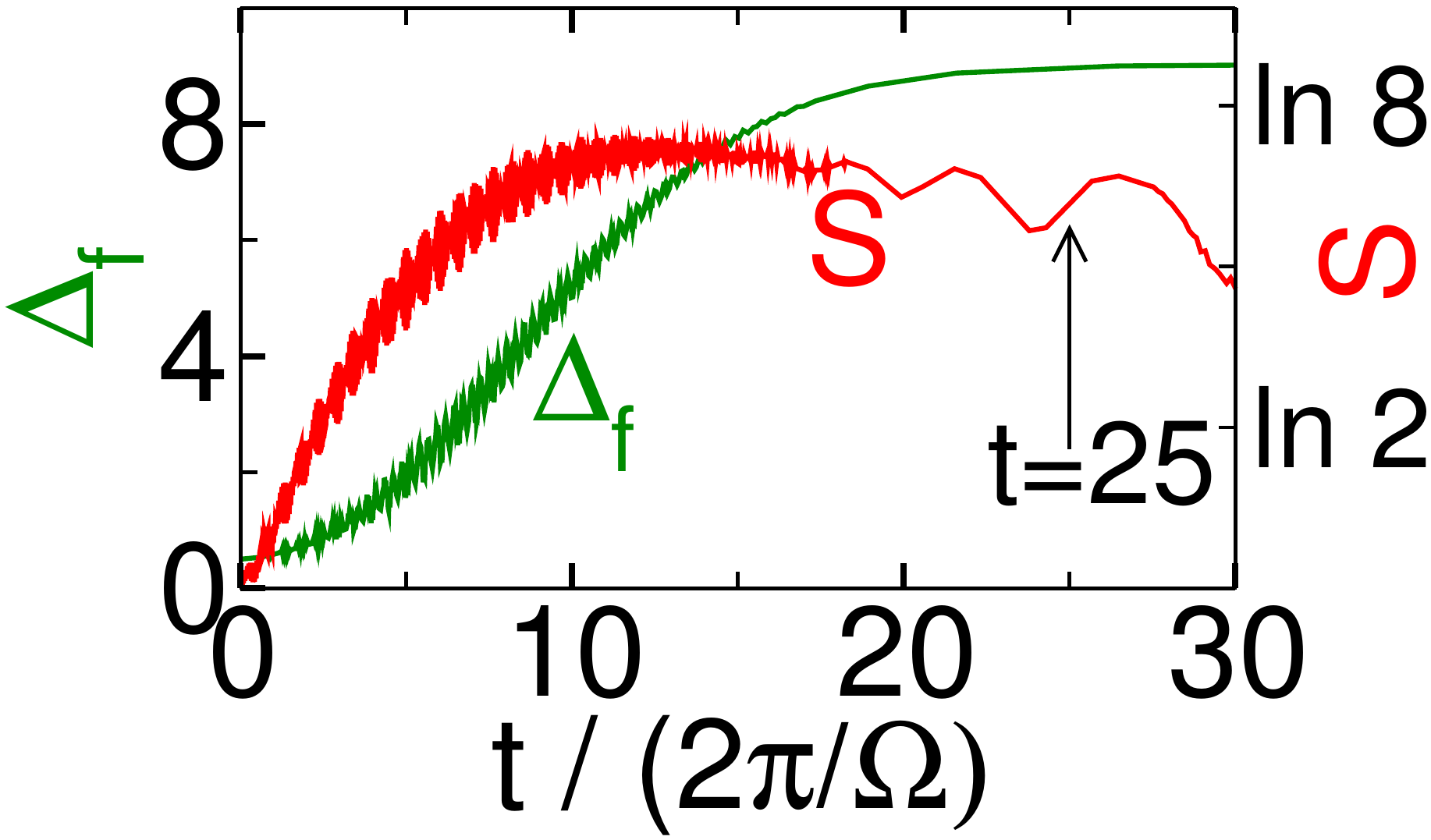} \\[1ex]
\includegraphics[width=0.33\linewidth]{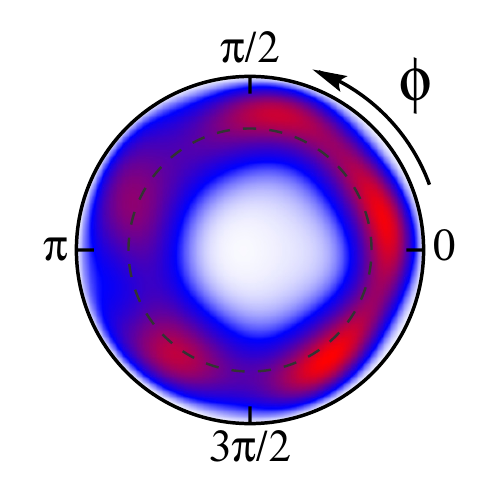} \hfill
\includegraphics[width=0.33\linewidth]{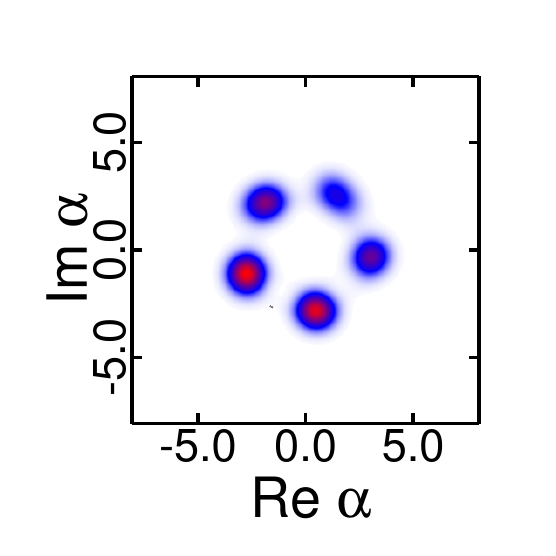} \hfill
\includegraphics[width=0.31\linewidth]{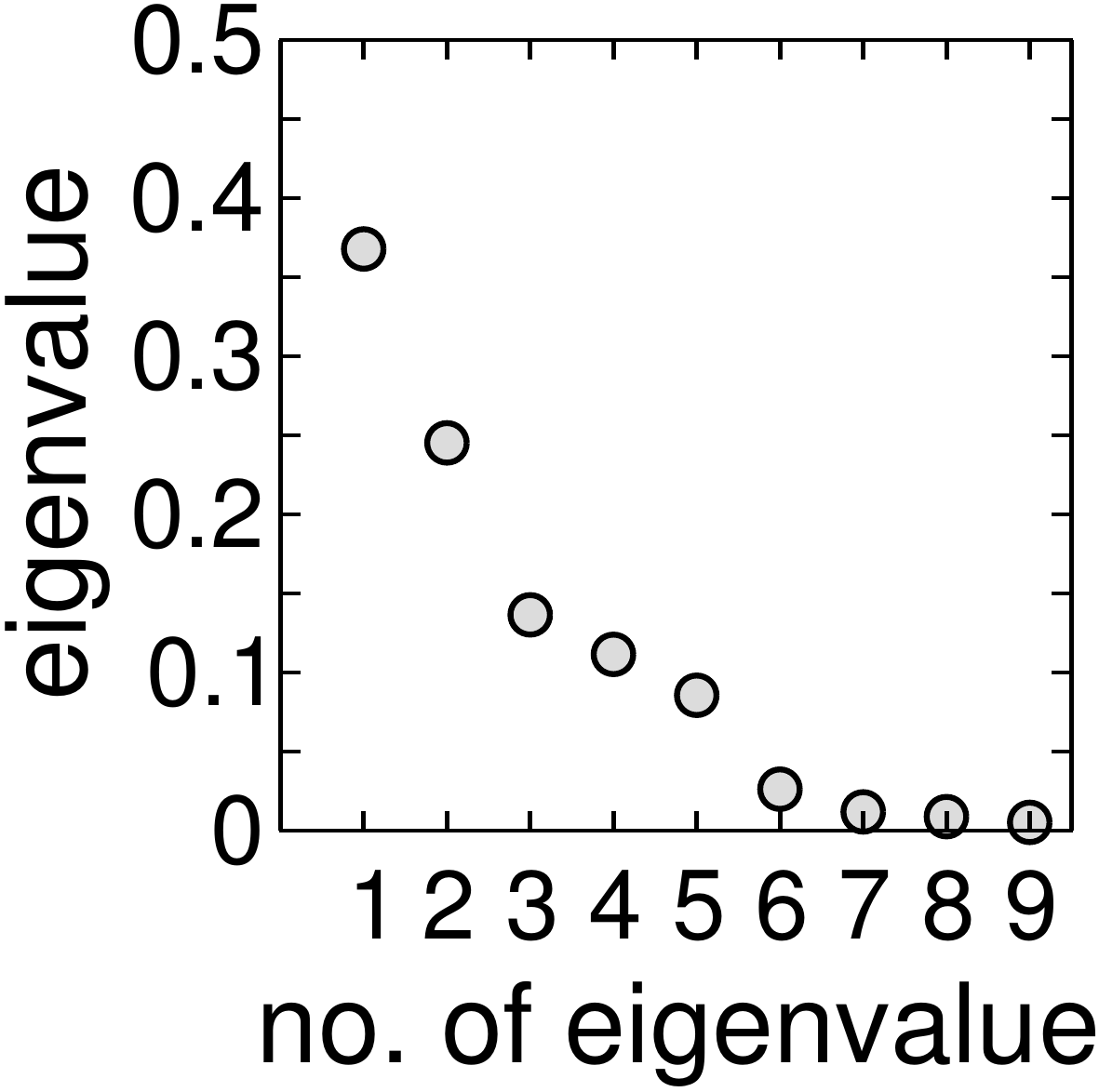} 
\caption{(Color online) Formation of field cat states in the CR dynamics for  $\Omega/\Delta=1/20 \ll\!1$, with $\lambda/\Delta=0.02$, $j=10$ 
and $\theta_0=\pi/4$, $\alpha_0=3$.
Upper panels: Spin expectation values $\langle J_{x/z} \rangle$ (left),
entanglement entropy $S$ and field variance  $\Delta_f$ (right).
Lower panels: 
Atomic and field Husimi function $Q_a(\theta,\phi)$ (left), $Q_f(\alpha)$ (middle)
and Schmidt coefficients (right), at $t=25 \times (2\pi/\Omega)$.
For $Q_a(\theta,\phi)$, we show the hemisphere $0 \le \theta \le \pi/2$,
with $\theta=0$ in the center, $\theta=\pi/2$ on the outer circle,
and $0 \le \phi < 2\pi$ running counterclockwise as depicted.
The gray dashed circles in this and the following figures indicate $\theta=\pi/4$.
}
\label{fig:NegDet}
\end{figure}

\section{Generation of atomic and field cat states}
\label{sec:Cat}

\begin{figure}[b]
\includegraphics[width=0.49\linewidth]{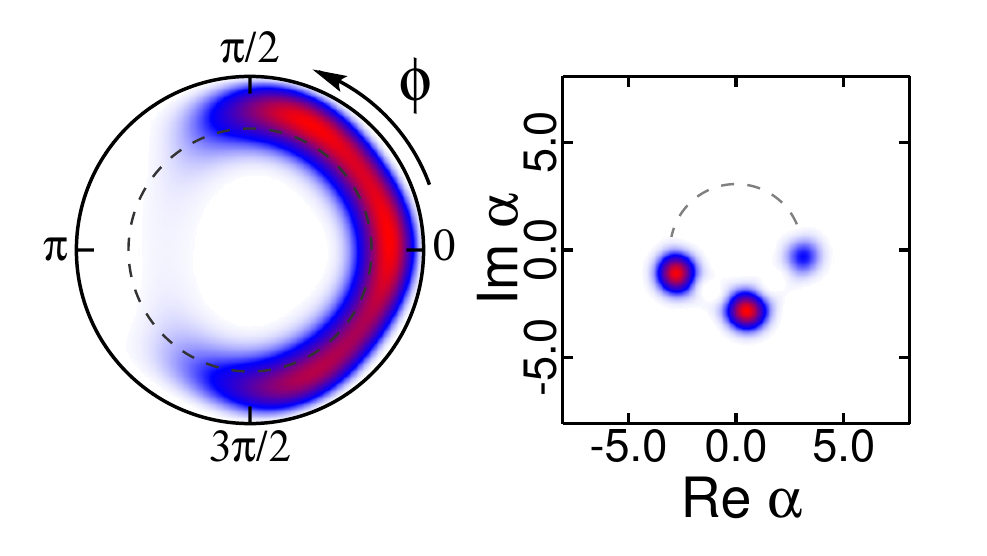} \hfill
\includegraphics[width=0.49\linewidth]{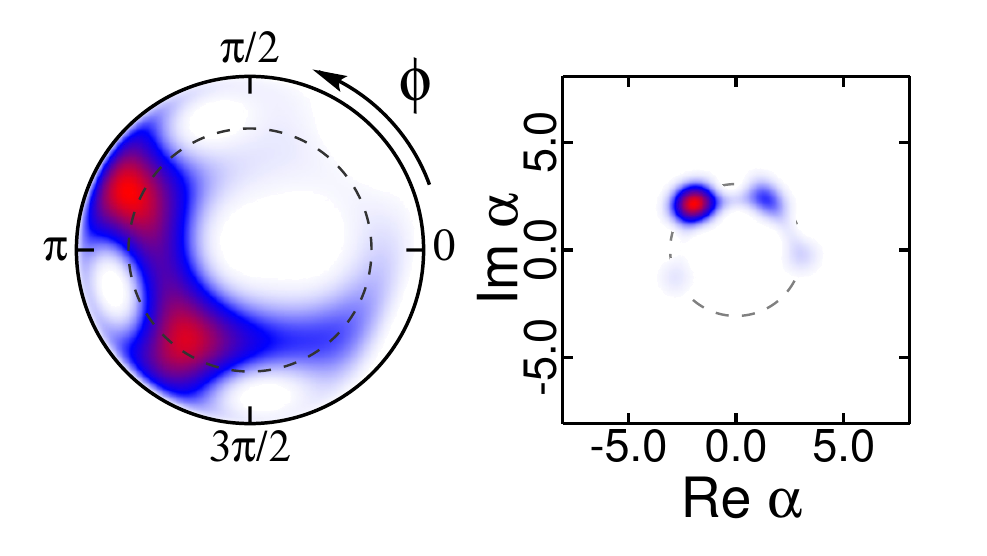}
\caption{(Color online) 
Atomic/field Husimi function of the two largest contributions in the Schmidt decomposition from Fig.~\ref{fig:NegDet},
with respective weight $0.37$ (left), $0.25$ (right).
}
\label{fig:NegDetDecomp}
\end{figure}

The non-resonant perturbation theory from the previous section shows that in the opposite cases $\Omega \ll \Delta$ and $\Omega \gg \Delta$ 
either the term $\propto a^\dagger a J_z $ or $\propto J_z^2$ in the effective Hamiltonian $\tilde{H}$ determines the dynamical properties over the respectively shorter time scale $T_E$ or $T_S$.
Both terms lead to a different structure of the wave function that 
is related to the generation of either field or atomic cat states.
These cat states appear as linear combinations of well-separated atomic or field coherent states.
A ``perfect'' field cat state would, e.g., be the state $|\alpha\rangle \pm |{-\alpha}\rangle$ for $|\alpha| \gg 1$.

We have discussed the case $\Omega \ll \Delta$ already in some detail,
and will now revisit the structure of the wave function for the situation shown in Fig.~\ref{fig:NegDet}.
As expected, the collapse phase coincides with large atom-field entanglement and large field variance. 
The coherent state parameters $\alpha_m(t)$ in the wave function Eq.~\eqref{PsiStruct} 
differ by an angle that is a multiple of $2\pi (t/T_E)$.
Whenever $t/T_E$ is a rational number, some of the $\alpha_m$ are equal such that fewer coherent field states appear in the wave function.
This explains the dips in the entanglement entropy $S(t)$ in Figs.~\ref{fig:CR1},~\ref{fig:NegDet}.

At $t=25 \times (2 \pi / \Omega) \approx  T_E/5$,
the field Husimi function $Q_f$ is a superposition of five Gaussian peaks.
For the atomic state, 
a similar fivefold pattern cannot be clearly identified in the atomic Husimi function~\cite{ZFG90} 
\begin{equation}
Q_a(\theta,\phi) = |\langle \theta,\phi|\psi\rangle|^2  \;, 
\end{equation}
which is defined via atomic coherent states from Eq.~\eqref{SpinCohGen} similar to the field Husimi function.
It gives the pseudo-spin probability distribution as a function of spherical angle coordinates $\theta, \phi$. For $\lambda=0$, the phase space sphere would rotate rigidly around the $z$-axis passing through the origin $\theta=0$.

\begin{figure}
\includegraphics[width=0.49\linewidth]{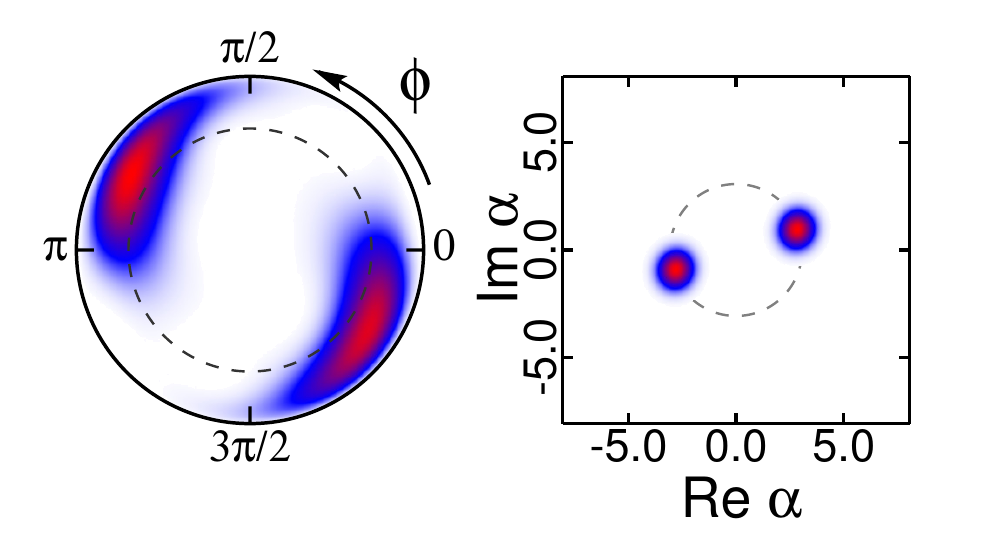} \hfill
\includegraphics[width=0.49\linewidth]{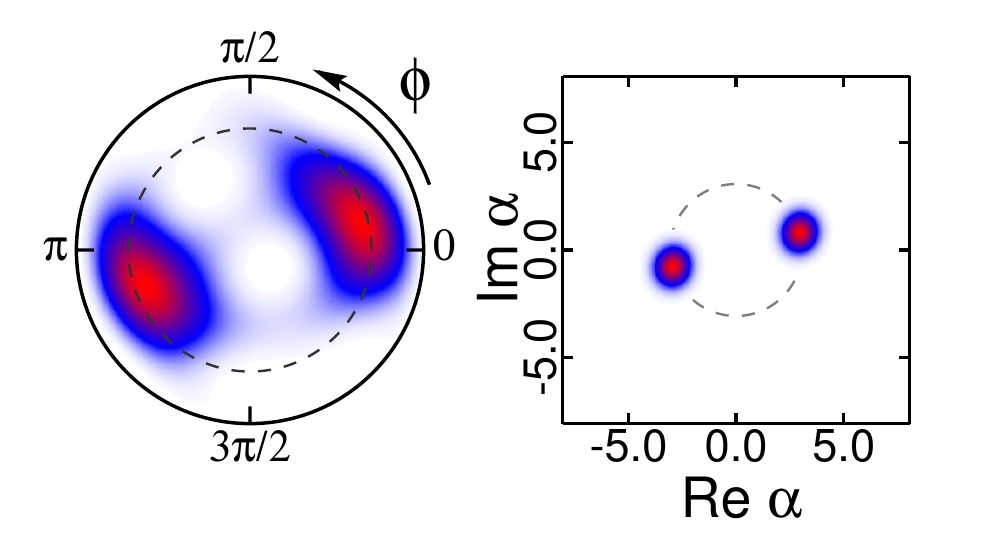}
\caption{(Color online) 
Atomic/field Husimi function of the two relevant contributions in the Schmidt decomposition 
at $t = 60 \times (2\pi/\Omega) \approx T_E/2$ (remaining parameters as in Fig.~\ref{fig:NegDet}),
with respective weight $0.65$ (left), $0.32$ (right).
}
\label{fig:NegDetHalfWay}
\end{figure}

\begin{figure}[b]
\includegraphics[width=0.48\linewidth]{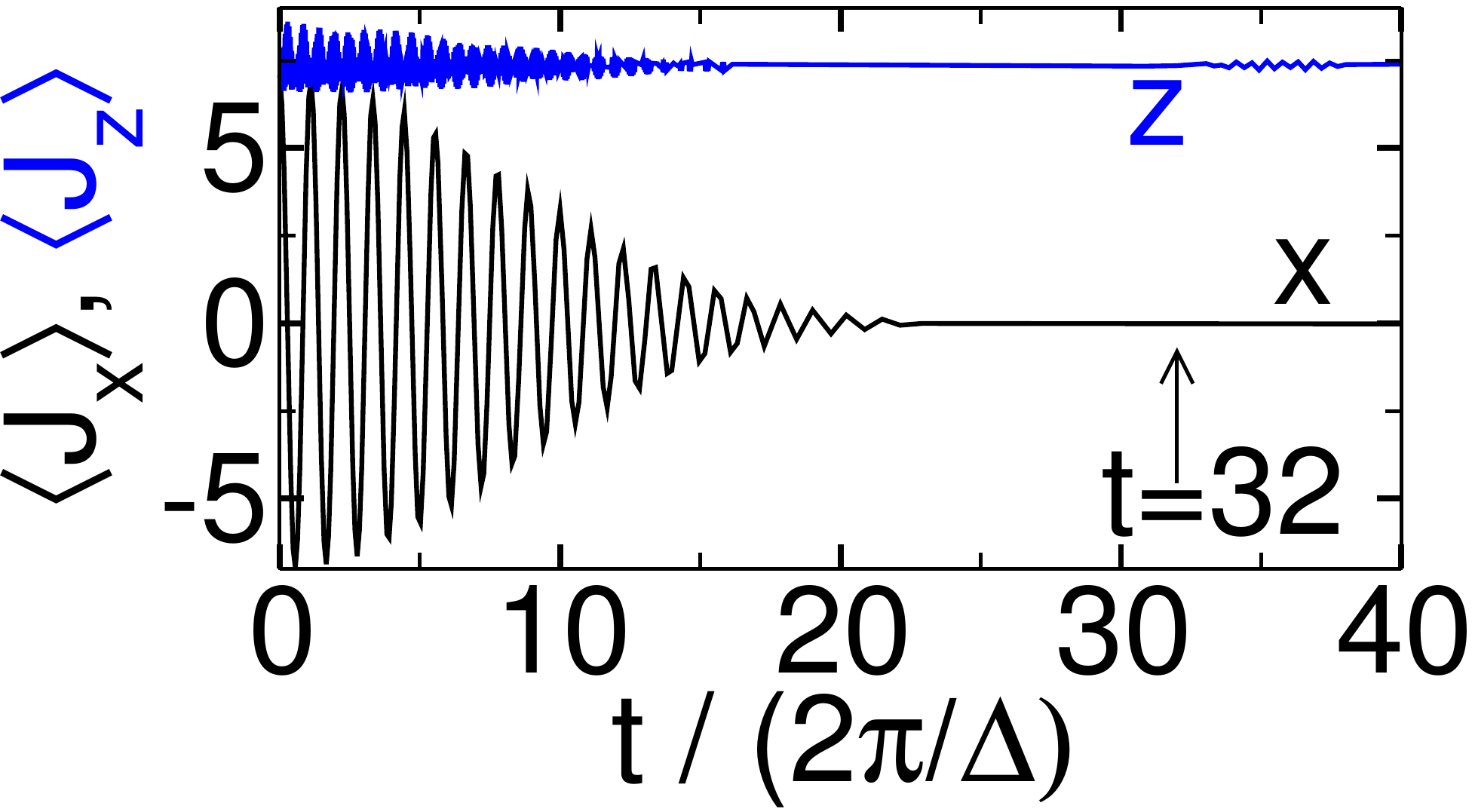} \hfill
\includegraphics[width=0.48\linewidth]{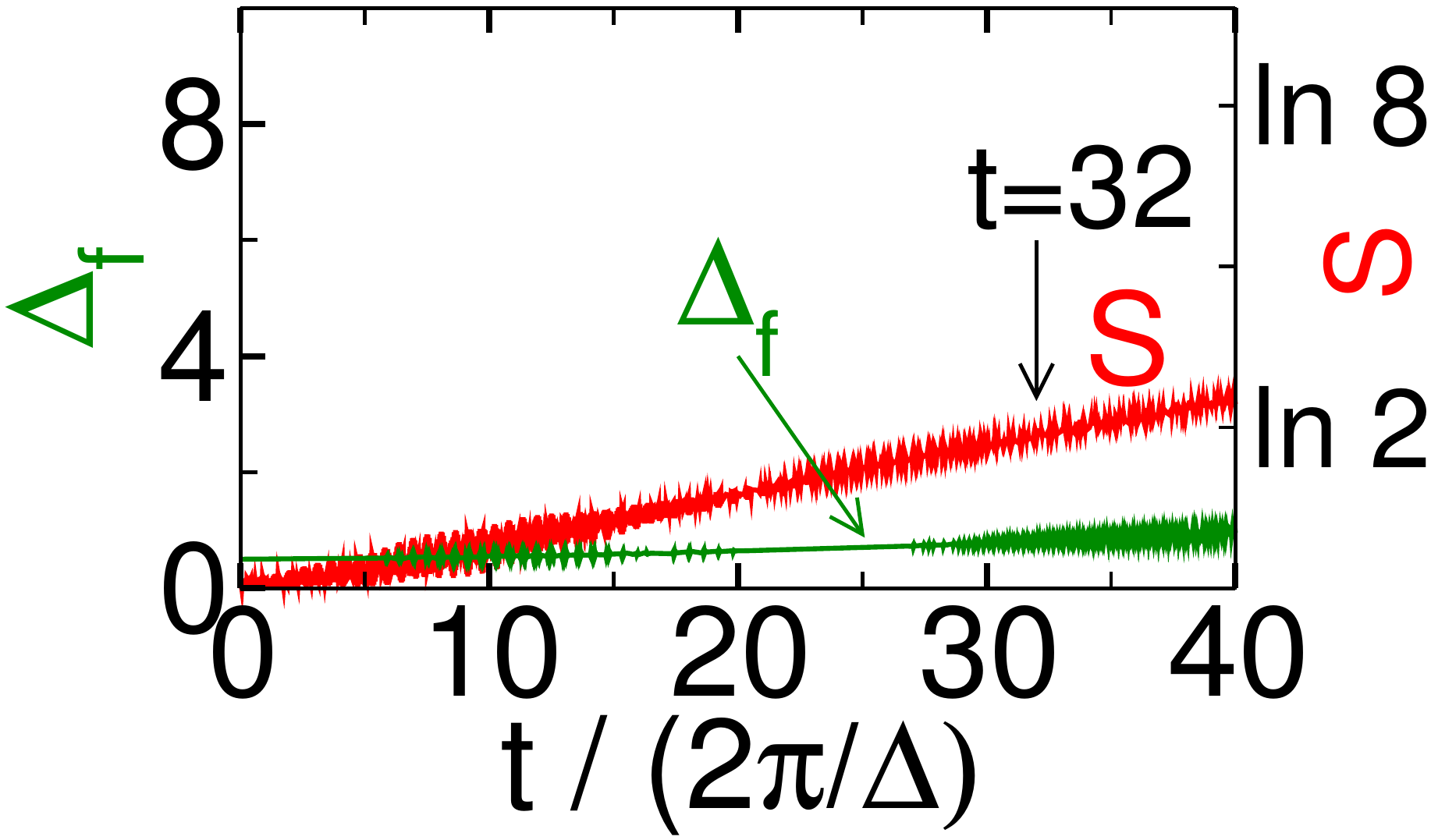} \\[1ex]
\includegraphics[width=0.33\linewidth]{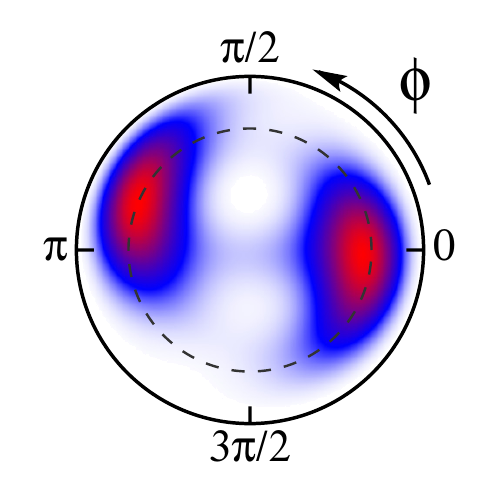} \hfill
\includegraphics[width=0.33\linewidth]{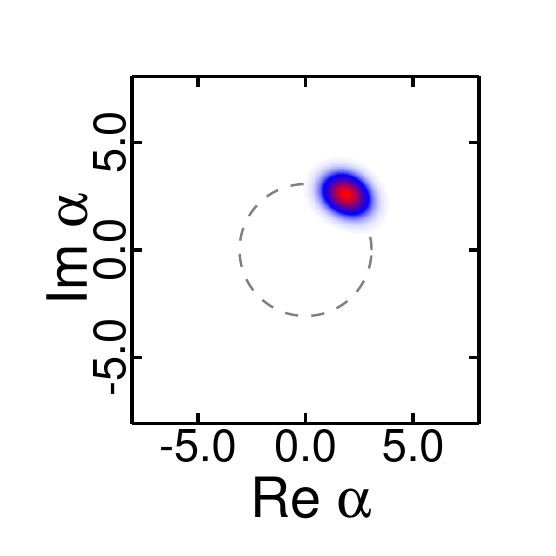} \hfill
\includegraphics[width=0.31\linewidth]{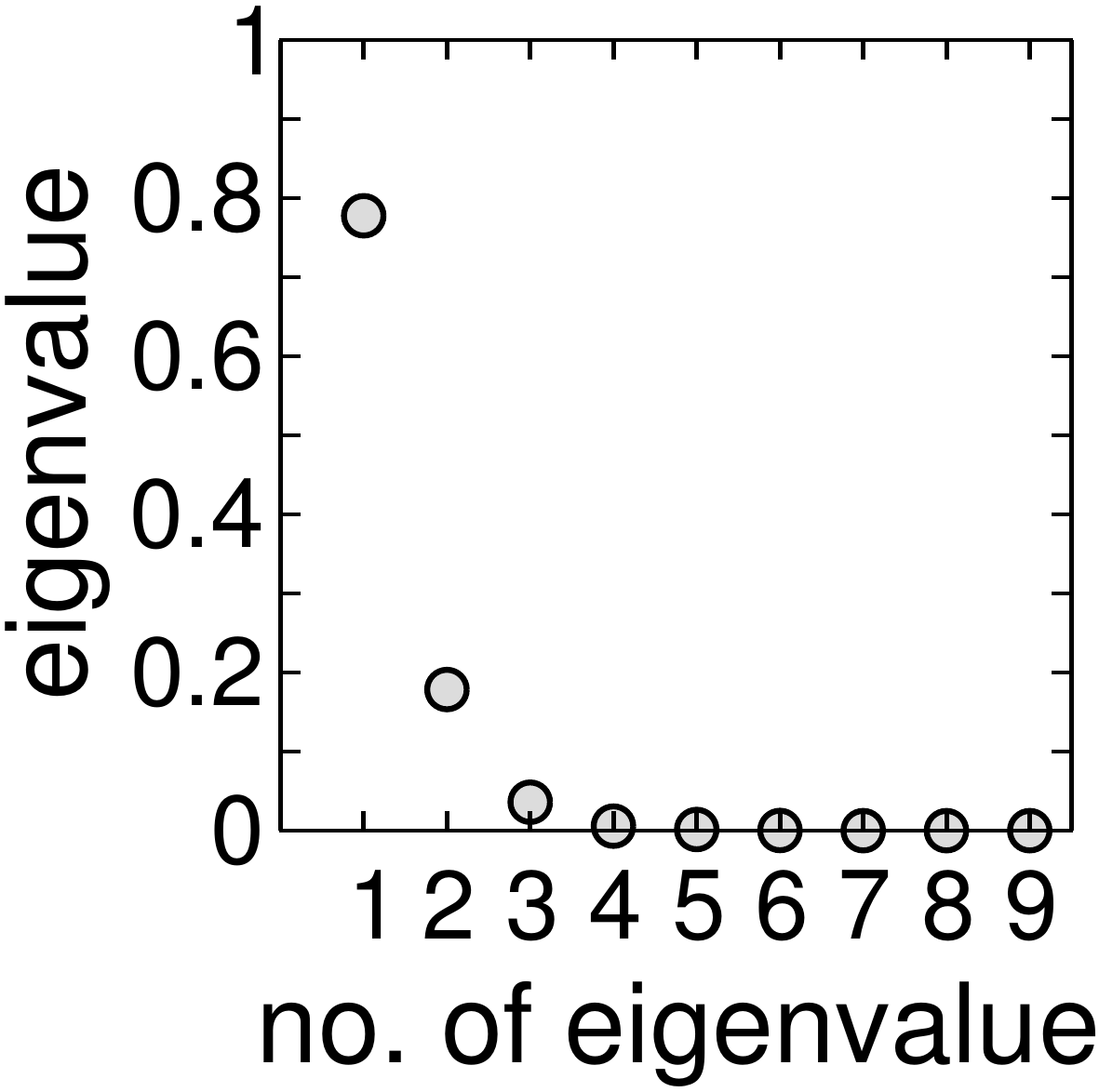} 
\caption{(Color online) Formation of atomic cat states for  $\Omega/\Delta=20 \gg\! 1$, with $\lambda/\Delta=0.5$, $j=10$ and $\theta_0=\pi/4$, $\alpha_0=3$ in the initial state.
The panels show the same quantities as in Fig.~\ref{fig:NegDet},
with $t=32 \times (2\pi/\Delta)$ in the lower row. }
\label{fig:PosDet}
\end{figure}

If the atomic states $|\sigma_m(t)\rangle$ in Eq.~\eqref{PsiStruct} were mutually orthogonal
the field superposition seen in $Q_f$ would be completely incoherent. 
We know from perturbation theory that the states are not orthogonal due to the rotation with $U^\dagger$ in Eq.~\eqref{SigmaM}, which allows for a coherent superposition and the appearance of field cat states.
To check up on this possibility we use the Schmidt decomposition of the atomic/field wave function~\cite{HHHH09}.
The Schmidt coefficients, i.e. the eigenvalues of both the atomic/field density matrix depicted by $Q_{a/f}$, have five relevant contributions.
The two largest are shown separately in Fig.~\ref{fig:NegDetDecomp}. 
We can identify a field cat state in the largest contribution (left panels), 
while the corresponding atomic state is strongly squeezed but not a cat state.
In the second largest contribution the indication of an atomic cat state is visible.
We also conclude that the fivefold field superposition in Fig.~\ref{fig:NegDet} 
 is partially coherent.

Halfway through the collapse phase, for $t \approx T_E/2$, the two contributions shown in Fig.~\ref{fig:NegDetHalfWay} comprise 98\% of the wave function.
Now, atomic and field cat states appear simultaneously.
Note that the appearance of field cat states is again a consequence of the transformation $U^\dagger$ in the perturbative result and the resulting non-orthogonality of the $|\sigma_m\rangle$ states in Eq.~\eqref{PsiStruct}.
Otherwise, the wave function would have the form $|\sigma_+\rangle |\alpha\rangle + |\sigma_-\rangle |{-\alpha}\rangle$ with two orthogonal atomic states $|\sigma_\pm\rangle$,
and no field cat states could appear.
Instead, the final rotation with $U^\dagger$ in Eq.~\eqref{SigmaM} leads to a finite overlap $\langle \sigma_+ |\sigma_-\rangle \ne 0$, and field cat states $ \simeq |\alpha\rangle \pm |{-\alpha}\rangle$ occur in the Schmidt decomposition.

\begin{figure}[t]
\includegraphics[width=0.49\linewidth]{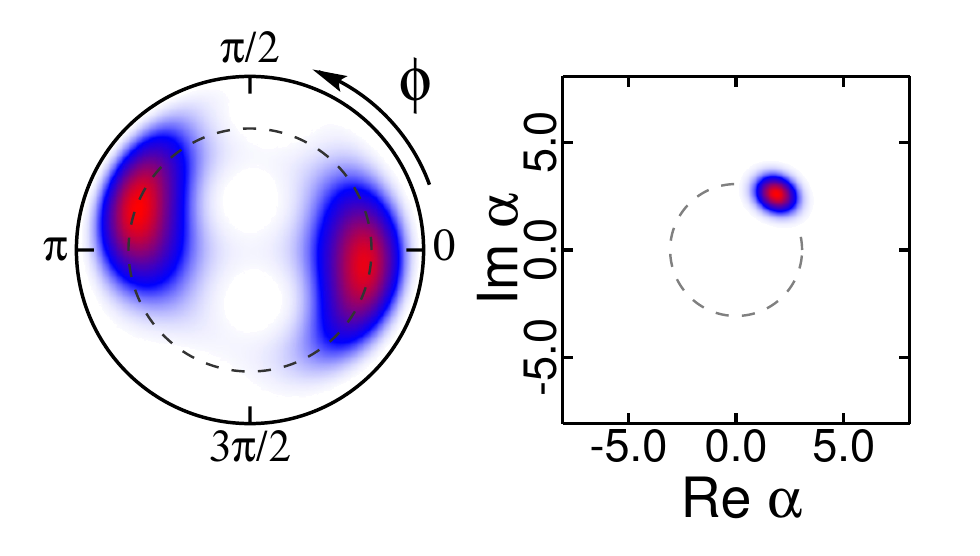} \hfill
\includegraphics[width=0.49\linewidth]{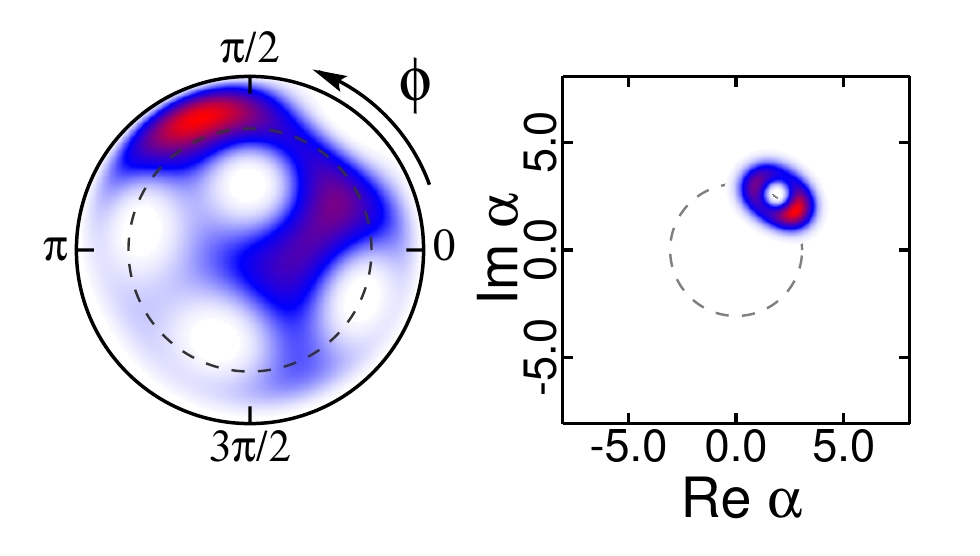}
\caption{(Color online) Atomic/field Husimi function of the two largest contributions in the Schmidt decomposition from Fig.~\ref{fig:PosDet},
with respective weight $0.78$ (left), $0.18$ (right).
}
\label{fig:PosDetDecomp}
\end{figure}

In the opposite case $\Omega \gg \Delta$, addressed in Fig.~\ref{fig:PosDet},
the wave function remains close to a product form $|\sigma(t)\rangle \otimes |\alpha(t)\rangle$
for times $t \lesssim T_S \ll T_E$.
The entanglement entropy and field variance are correspondingly small.
In contrast to the initial short-time dynamics of Rabi oscillations in Eq.~\eqref{PertRabi}  that we deduced from the product form for times $t \ll T_S, T_E$, the term $J_z^2$ in $\tilde{H}$ now gives rise to squeezing and splitting of the initial atomic coherent state.
In the situation shown in Fig.~\ref{fig:PosDet}, 
the atomic $Q$-function at $t=32 \times (2\pi/\Delta) \approx  T_S/2 \ll T_E$ 
has a two-fold structure.
It appears since,  in the present case of integer even $j$,
\begin{equation}
  \exp\Big[-\ii \frac{\pi }{2} J_z^2 \Big] |\theta\rangle =  \frac{1}{1+\ii}  |\theta \rangle + \frac{1}{1-\ii} |{-\theta}\rangle 
\end{equation}
is a linear combination
of two atomic coherent states $|\theta \rangle$, $|{-\theta}\rangle$.
The two relevant contributions in the Schmidt decomposition are shown in Fig.~\ref{fig:PosDetDecomp}.
As opposed to the previous case $\Omega \ll \Delta$,
we identify the signatures of an atomic cat state together with a single coherent field state in the largest contribution (left panels).
The annulus in $Q_f$ for the second largest contribution
(right panels) originates from a superposition of two field coherent states $|{\alpha \pm \delta \alpha}\rangle$ with $|\delta \alpha| \ll 1$. 
This structure is a precursor of the field state splitting through the $a^\dagger a J_z$-term for later times $T_S \lesssim t \lesssim T_E$.

\section{Atom-field entanglement and the semi-classical approximation}
\label{sec:SCA}

The CR mechanism for $\Omega \ll \Delta$ 
does not depend on the non-classical properties of a quantized field,
but only on the possibility of atom-field entanglement.
We thus expect that the basic signature of this mechanism, the collapse of Rabi oscillations,
occurs also for atoms in strong radiation fields close to the classical field limit
where the field quantization plays no role.

The classical field limit can be defined rigorously as the limit
$|\alpha_0| \to \infty$, keeping $\lambda |\alpha_0|$ constant.
In this limit, the field mode evolves independently of the atoms because the product $\lambda \langle J_x \rangle$, which gives the strength of the atomic influence on the field, goes to zero for $\lambdaÊ\to 0$ and finite $j$.
The product $\lambda \langle a + a^\dagger \rangle$, which determines the influence of the field on the atoms, remains finite.
The Dicke model reduces to the model of an atomic ensemble driven by an external field
$\mathbf{B}(t) = ( 2 \lambda \alpha_0 \cos \Omega t, 0, - \Delta)$.
The atomic expectation values $\mathbf{J}=(\langle J_x \rangle, \langle J_y \rangle, \langle J_z \rangle)$ obey the equation of motion
\begin{equation}\label{JDriv}
 \partial_t \mathbf{J} = \mathbf{B}(t) \times \mathbf{J} \;.
\end{equation}
The characteristic signatures of such a driven system are Rabi oscillations.

\begin{figure}
\includegraphics[width=0.46\linewidth]{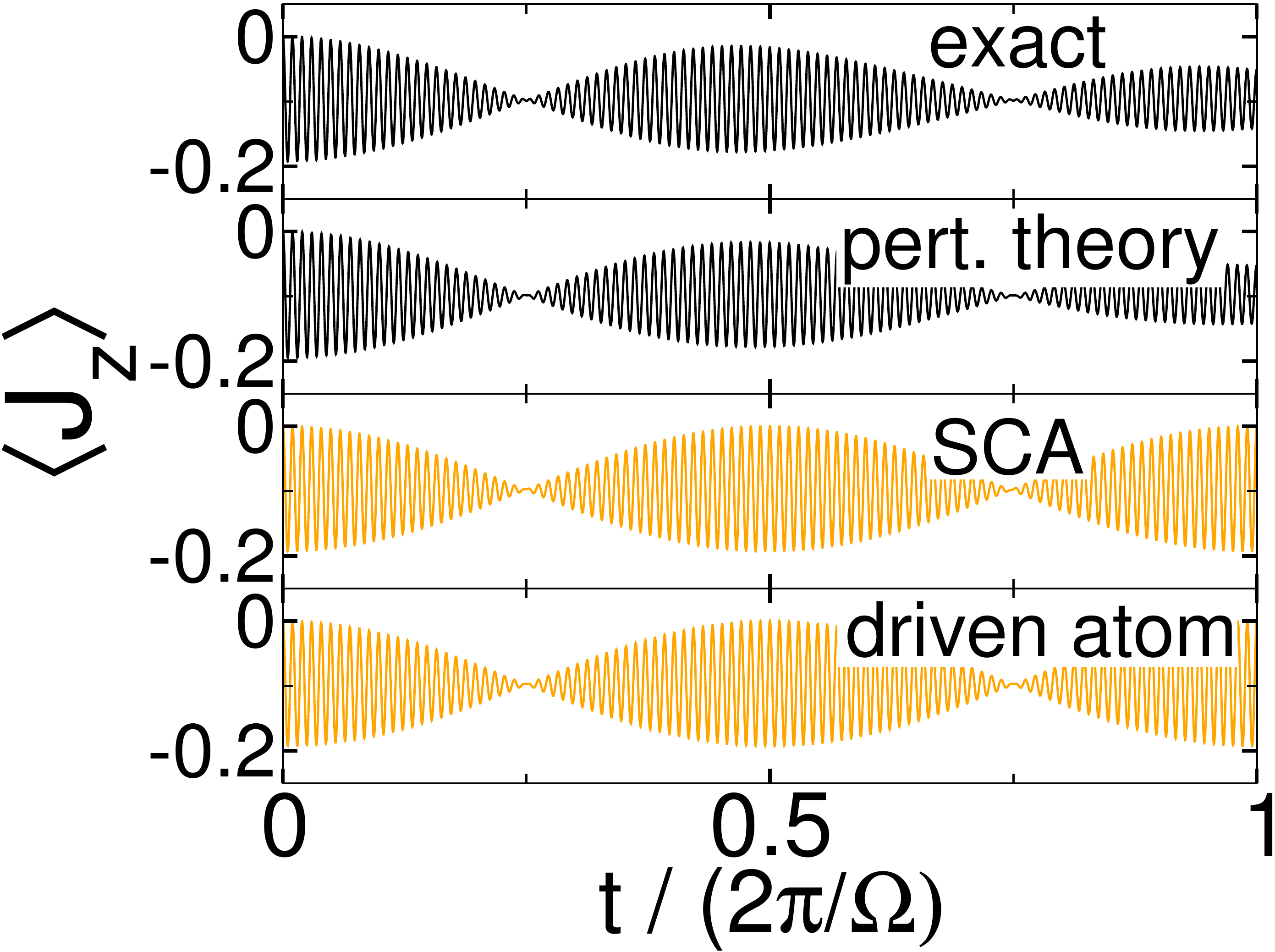} \hfill
\includegraphics[width=0.49\linewidth]{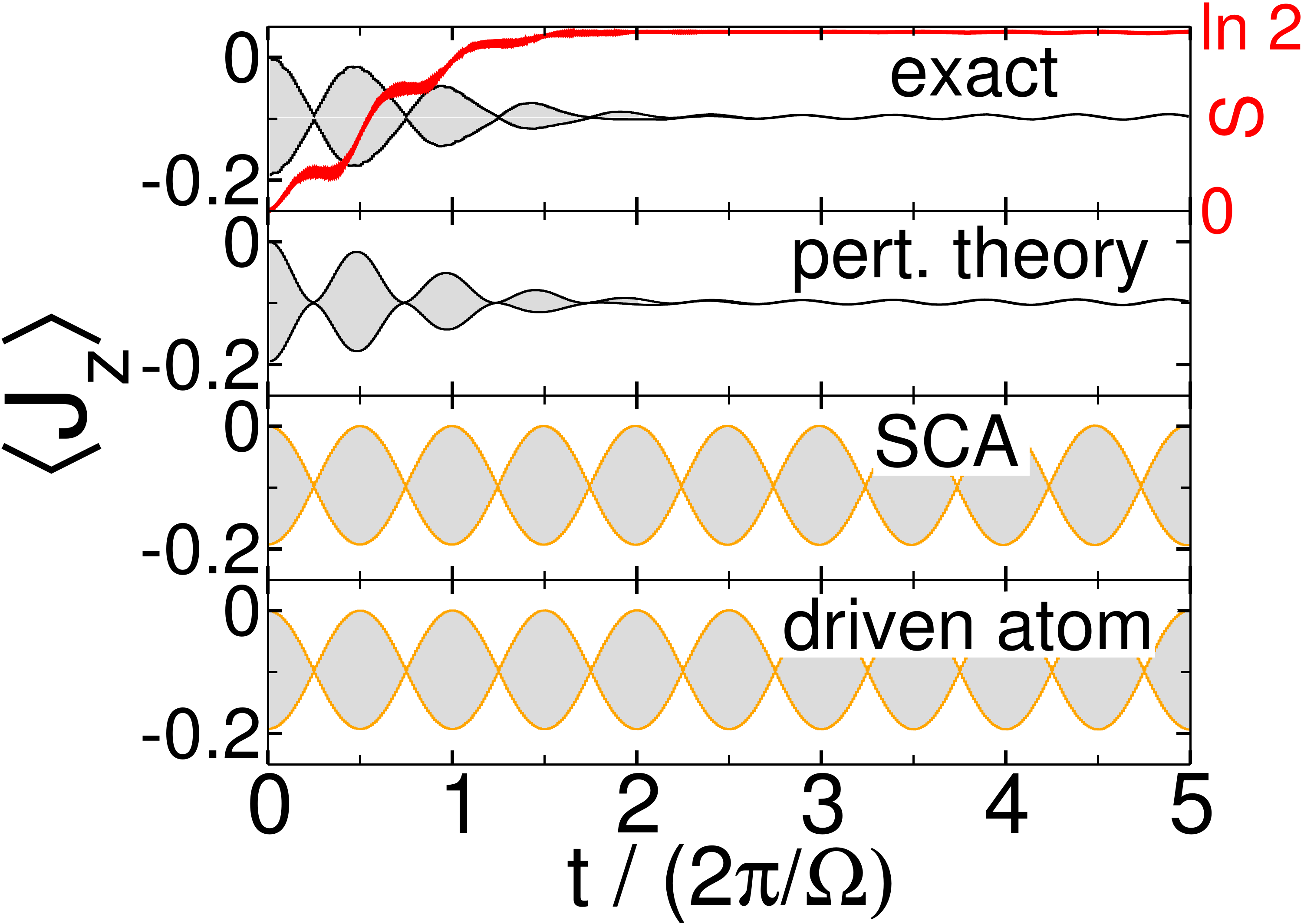}
\caption{(Color online) Comparison 
of the SCA to the exact dynamics of the Rabi model ($j=1/2$) with $\Omega/\Delta=0.01$, 
$\lambda/\Delta=0.02$ and $\theta_0=\pi/2$, $\alpha_0=5$.
The left panels show $\langle J_z(t) \rangle$ over a short time interval,
presenting the exact numerical, perturbative and SCA result together with the result in the limit of a field-driven atom (Eq.~\eqref{JDriv}). The right panels cover a longer time interval and include the entanglement entropy $S(t)$ (red curve) from the numerical calculation.
}
\label{fig:SCA}
\end{figure}

For finite $\alpha, \lambda$ 
the field mode does not evolve independently of the atoms,
but we expect only small corrections from the classical field dynamics for small $\lambda$.
One attempt to include these corrections is the semi-classical approximation (SCA) (see, e.g., Ref.~\cite{GH84} for a discussion).
The SCA is based on the assumption that the coupled atom-field system
remains in a product $|\theta(t), \phi(t) \rangle \otimes |\alpha(t) \rangle$
of coherent states during time evolution.
This assumption allows for decoupling of the equations of motion
for the atomic and field expectation values.

After decoupling, the atomic state evolves again similar to a spin in a magnetic field
$\mathbf{B}(\alpha) =( 2 \lambda \Re \alpha, 0, - \Delta)$.
Now, however, the field state evolves as for an oscillator with an additional external force $\lambda \langle J_x \rangle$ that accounts for the back-reaction from the atomic ensemble.
The corresponding SCA equations of motion are
\begin{equation}\label{SCA}
 \partial_t \mathbf{J} = \mathbf{B}(\alpha) \times \mathbf{J} \;,
 \quad  \ii \partial_t \, \alpha = \Omega \alpha + \lambda \langle J_x \rangle   \;.
\end{equation}
The SCA equations of motion become exact in the classical field limit,
where they reduce to Eq.~\eqref{JDriv}.

From the equations of motions we can observe a potential problem of the SCA that arises from the generation of atom-field entanglement in the true Dicke dynamics.
The change of the field state in Eq.~\eqref{SCA} is $\propto \lambda$,
and the influence on the atoms $\propto \lambda^2$.
Therefore, the dynamically relevant time scale in SCA is $ \propto 1/\lambda^2$, 
just as the scaling of the entangling time $T_E$ from Eq.~\eqref{TS}.
We should thus expect that for time scales on which the SCA differs from the simpler Eq.~\eqref{JDriv} significant atom-field entanglement has been generated for which the SCA cannot account.

We consider exemplarily the limit $\Omega \to 0$ of a classical field with negligible energy quantization. 
In Fig.~\ref{fig:SCA} we compare the SCA with the exact and perturbative result for the dynamics,
and with the simplified Eq.~\eqref{JDriv} of a driven atomic ensemble.
For short times (left panels) all four descriptions agree and describe Rabi oscillations,
which are characteristic for classical field dynamics.
For longer times (right panels), we see that significant entanglement is generated over the first few field periods even in the most simple Rabi case $j=1/2$.
The SCA cannot account for entanglement and accordingly misses the collapse of Rabi oscillations  entirely.  We note that the SCA simply reproduces the result also obtained with the simpler Eq.~\eqref{JDriv}.
Note also that we are in the weak coupling regime, where the non-resonant perturbation theory describes the dynamics accurately.
The failure of the SCA is not a result of strong atom-field coupling.

\begin{figure} 
\includegraphics[width=0.47\linewidth]{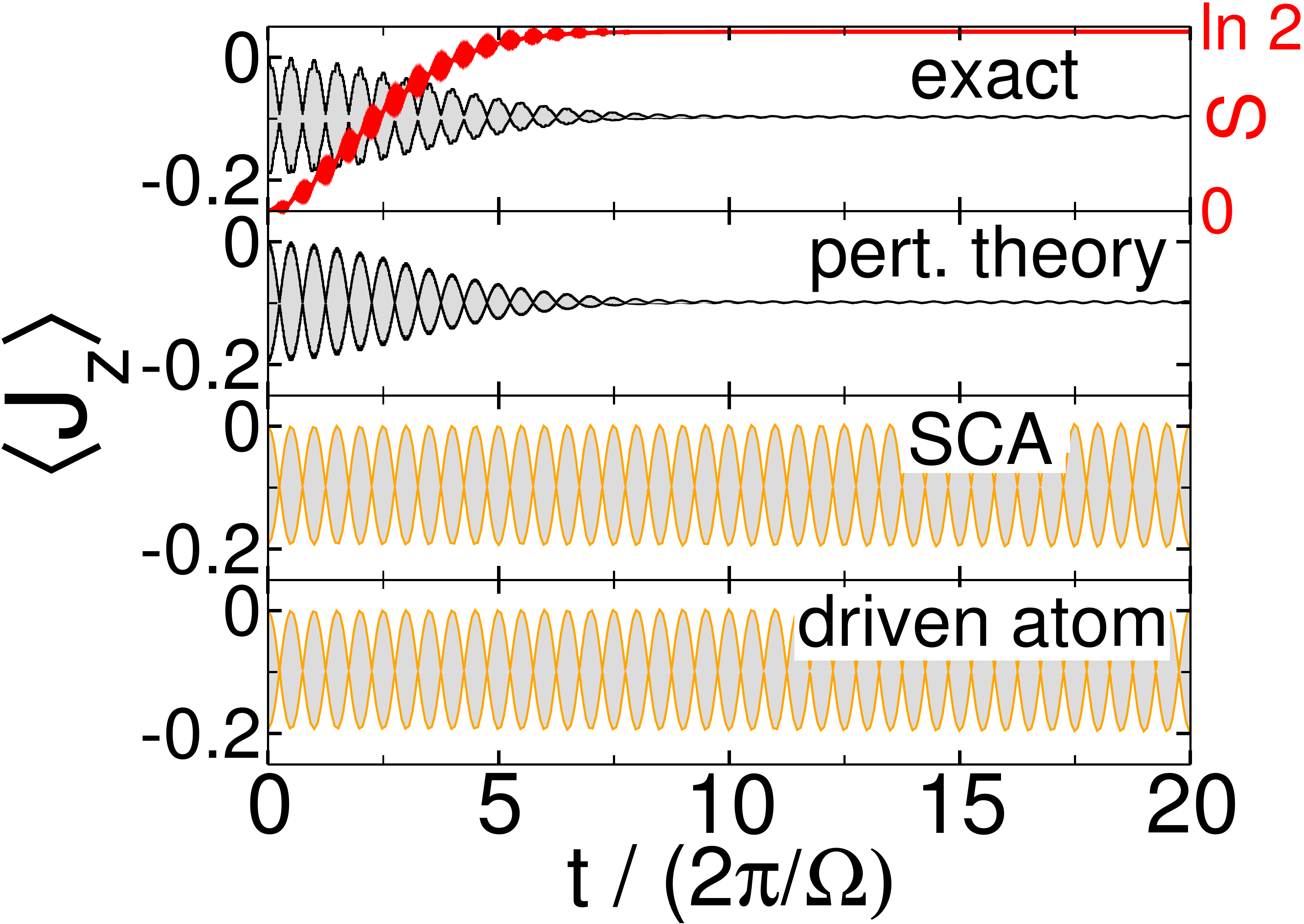} \hfill
\includegraphics[width=0.48\linewidth]{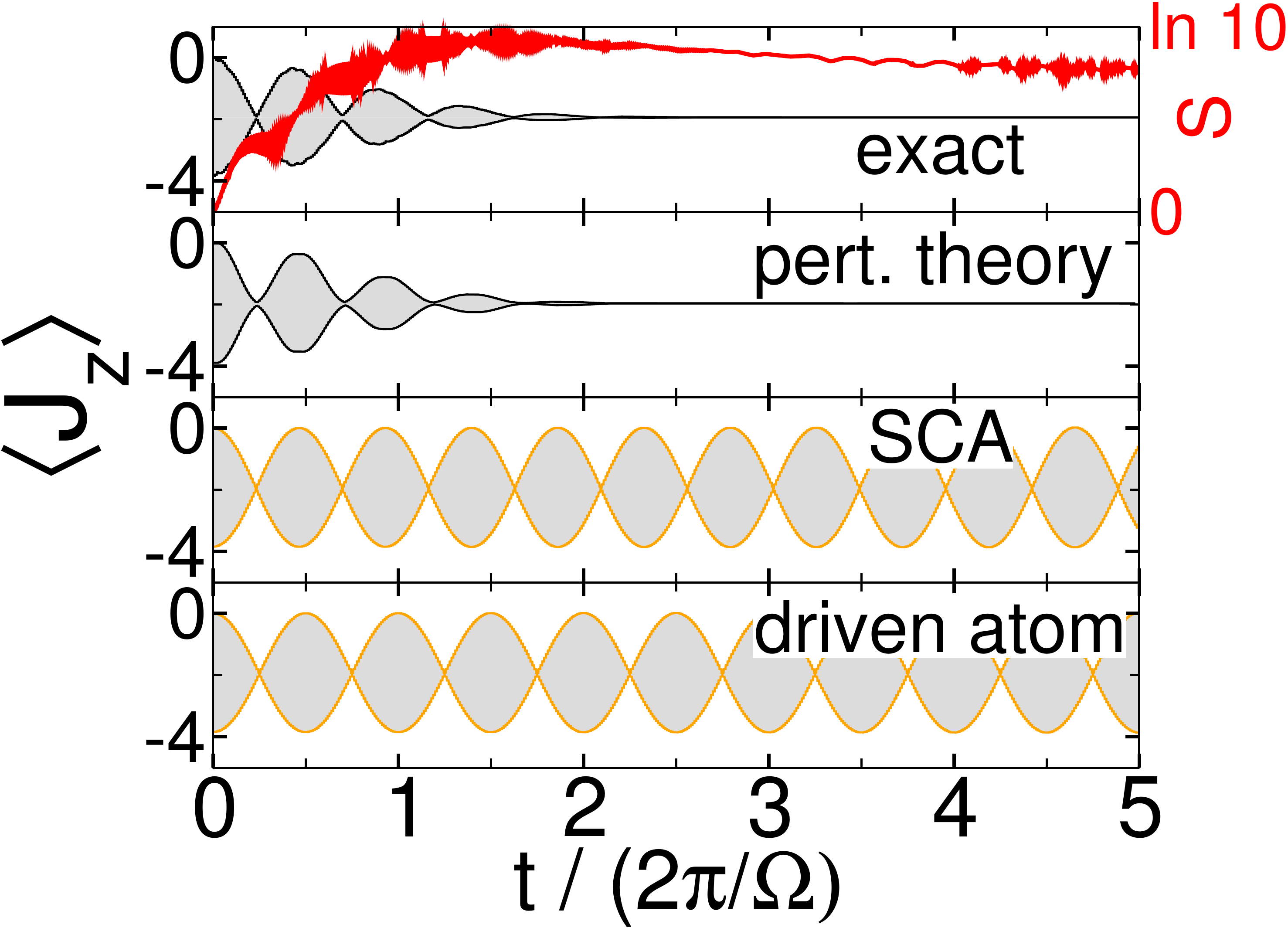}
\caption{(Color online) Comparison 
of the SCA to the exact dynamics, still with $\Omega/\Delta=0.01$ and  $\theta_0=\pi/2$ as in the previous Fig.~\ref{fig:SCA}.
Left panels: Results for the Rabi model ($j=1/2$) with $\lambda/\Delta=0.005$ and $\alpha_0=20$.
Right panels: Results for the Dicke model with $j=10$, and other parameters
$\lambda/\Delta=0.02$ and $\theta_0=\pi/2$, $\alpha_0=5$ identical to Fig.~\ref{fig:SCA}.
}
\label{fig:SCA2}
\end{figure}

A situation with larger $\alpha_0$ is shown in the left panel in Fig.~\ref{fig:SCA2}.
The collapse of Rabi oscillations takes place for later times since $T_E$ is larger,
but both SCA and the simpler Eq.~\eqref{JDriv} again fail in the same way.
In the right panel in Fig.~\ref{fig:SCA2} we see that this behavior is not restricted to the Rabi case,
but occurs equally for $j \gg 1/2$.
We note that its significance is not reduced 
since the present violation of classical field dynamics is not a consequence 
of quantum fluctuations of the atomic system, which would decay with $1/j$.
The relevance of entanglement, bounded by $\ln (2j+1)$, even increases with $j$.

The violation of the mean-field assumption of negligible atom-field correlations already at weak coupling presents a genuine problem for the SCA.
As a consequence of entanglement generation, the SCA can be correct only as long as the field dynamics remains decoupled from the dynamics of the atomic ensemble.
In that situation the atomic ensemble is already described by the simpler Eq.~\eqref{JDriv}:
The SCA does not improve on a model with entirely classical field dynamics where any influence of the atoms on the field state is neglected.
We conclude that the SCA fails to account for the back-reaction of the atoms on the field,
even at weak coupling or large $j$. 

The non-resonant CR dynamics discussed here is an example of non-trivial coupled quantum-classical dynamics~\cite{DGS00}, as is evident from the wave function Eq.~\eqref{PsiStruct} with multiple coherent field states. In our case, the signatures of coupled quantum-classical dynamics are the
periodic CR patterns which cannot be explained in a simple mean-field description.

\section{Conclusions}
\label{sec:Conc}

In summary, our analysis explains the CR patterns of Rabi oscillations in the non-resonant Dicke model by the dynamical splitting of atomic or field coherent states,
which takes place on different time scales distinguished by the detuning ratio $\Omega/\Delta$.
The collapse phase is accompanied by the buildup of atom-field entanglement or atomic squeezing,
whose subsequent decay leads to periodic revivals.
Both atomic and field cat states can arise in the collapse phase.
The quantitative explanation of this behavior is provided by perturbation theory for large detuning.
Application of the RWA to the non-resonant case would erroneously predict a single time scale (cf. App.~\ref{sec:app}), instead of the two time scales obtained in the correct calculation.
 
The non-resonant CR patterns arise through
a dynamical mechanism that involves the creation of 
highly non-classical states from initial classical state preparations.
They give direct evidence for quantum entanglement and coherent quantum superpositions in atom-field and related systems.
Even close to the classical field limit, where other quantum properties such as field quantization are of minor importance, atom-field entanglement can prevail over the semi-classical dynamics that would occur for a hypothetical non-correlated atom-field system.
This indicates how semi-classical approximations
can fail because they neglect atom-field correlations.

The observation of non-resonant CR patterns might become possible in experiments using superconducting circuits instead of optical cavities~\cite{SG08}.
These experiments can reach the regime of strong coupling or large detuning~\cite{AANPT08}, which requires calculations beyond the RWA~\cite{BGAANB09,ZRKH09}.
Although the experimentally controllable detuning can be made large,
the most serious obstruction against observation of the dynamical patterns described here  
is the necessity of preserving quantum coherence over many Rabi oscillations.
Further improvement of experimental techniques may resolve this issue.

\begin{acknowledgments}
This work was supported by 
DFG
through  AL1317/1-1 and SFB 652.
\end{acknowledgments}


\appendix

\section{Non-resonant perturbation theory for Dicke-type Hamiltonians}
\label{sec:app}

We give here the result of second order perturbation theory for Dicke-type Hamiltonians
\begin{equation}\label{app:H}
\begin{split}
  H = & \; \omega J_z + \Omega a^\dagger a \\
   &+ g ( a J_+ + a^\dagger J_-   )   + \bar{g} (a^\dagger J_+ + a J_-) \;,
  \end{split}
\end{equation}
with $J_\pm = J_x \pm \ii J_y$ denoting spin ladder operators,
in the  non-resonant case $|\omega| \ne \Omega$.

Standard perturbation theory gives a correction to the eigenstates and eigenvalues of the non-interacting Hamiltonian in the first line in Eq.~\eqref{app:H}.
The result can be expressed in the form of Eq.~\eqref{PertWave},
with a unitary transformation
\begin{equation}
  U = \exp[  T_1 + T_2  ] 
\end{equation}
that accounts for the change of the eigenstates,
where
\begin{equation}
 T_1 = \frac{g}{\omega-\Omega} (  a J_+ - a^\dagger  J_-  ) + \frac{\bar{g}}{\omega+\Omega}( a^\dagger J_+  -  a J_- ) \;,
\end{equation}
\begin{equation}
 T_2 = \frac{g \bar{g}}{ \omega^2 - \Omega^2 } \Big[  \frac{\Omega}{2 \omega} \big(J_+^2 - J_-^2 \big) 
  + \frac{\omega}{\Omega } \big( {a^\dagger}^2 - a^2 \big) J_z  \Big] \;,
\end{equation} 
and an effective Hamiltonian
\begin{equation}
\begin{split}
 \tilde{H} = & \; \omega J_z + \Omega a^\dagger a \, \\ 
 & +  \frac{g^2 (\omega + \Omega) + \bar{g}^2(\omega-\Omega) }{\omega^2 - \Omega^2}  (2 a^\dagger a + 1) J_z \\
 & + \frac{g^2 (\omega + \Omega) - \bar{g}^2(\omega-\Omega) }{\omega^2 - \Omega^2} \big( J^2 - J_z^2 \big)
  \end{split}
\end{equation}
that gives the perturbed eigenvalues.
By construction, $\tilde{H}$ is diagonal in the unperturbed eigenstates and contains only operators $J_z$ and $a^\dagger a$, and the constant of motion $J^2$.
Perturbation theory thus provides us with an approximate diagonalization of the Hamiltonian, in the sense that both sides of 
\begin{equation}\label{app:Cons}
 U^\dagger \tilde{H} U = H + O(\{g,\bar{g}\}^3)
\end{equation}
differ by terms of third or higher order in the coupling constants.
Alternatively, the contribution of $T_2$ can be included in the effective Hamiltonian as in Ref.~\cite{ZRKH09}.
The present formulation with a diagonal $\tilde{H}$ is preferential for the study of the dynamical evolution of the wave function since it allows for direct evaluation of Eq.~\eqref{PertWave}.

Our Hamiltonian from Eq.~\eqref{Ham} corresponds to the choice $g=\bar{g}$, including counter-rotating terms in $H$.
The second order term $T_2$ is finite and contributes to the atomic and field squeezing,
but it appears as a higher order correction to the leading order term $T_1$ in $U$.
Therefore, it is not relevant for the understanding of the dynamical effect discussed in the main part of the text and was not included in Eq.~\eqref{Utraf}.
Within the perturbative setting the relevant second order term in $\tilde{H}$, which grows during time evolution, is separated from the irrelevant second order term in $U$.

In the RWA, applicable at resonance $\omega \approx \Omega$,
counter-rotating terms are dropped from the Hamiltonian by setting $\bar{g}=0$.
In Refs.~\cite{APS97,KS98}
an even simpler model with $\bar{g}=0$ and $\Omega=0$ was taken as the starting point.
Within RWA, the term $T_2$ is zero in accordance with the fact that the number of excitations
(corresponding to the operator $J_z + a^\dagger a$) is conserved.
More importantly, the effective Hamiltonian now contains the same prefactor $g^2/(\omega-\Omega)$ in front of the two operators $a^\dagger a J_z$ and $J_z^2$.
Instead of the two different time scales for atomic and field squeezing introduced in Eq.~\eqref{TS}
only a single time scale appears in the RWA.
By construction, the RWA is valid only close to resonance and incompatible with the  
non-resonant perturbation theory.


%

\end{document}